\documentclass[reprint, onecolumn, amsmath, amssymb,superscriptaddress]{revtex4-2}
\usepackage{booktabs}
\usepackage{color, xcolor, colortbl}
\usepackage{graphicx,epstopdf}
\usepackage{geometry}
\usepackage{amsmath,amssymb,amsthm,amsfonts}
\usepackage{algorithm}
\usepackage{algorithmic}
\usepackage{dcolumn}
\usepackage{epstopdf}
\usepackage{bm}
\usepackage{subcaption}
\usepackage{array}
\usepackage{appendix}
\usepackage{multirow}
\usepackage{braket}
\usepackage[english]{babel}
\usepackage{hyperref}
\usepackage[capitalize]{cleveref}
\usepackage{xpatch}
\usepackage{tikz}
\usepackage{xspace}
\usepackage[roman]{complexity}
\usepackage{url}

\usepackage[export]{adjustbox}

\newcolumntype{M}[1]{>{\centering\arraybackslash}m{#1}}

\newcommand{\bvec}[1]{\mathbf{#1}}
\renewcommand{\vec}[1]{\bvec{#1}}

\makeatletter
\newcommand\footnoteref[1]{\protected@xdef\@thefnmark{\ref{#1}}\@footnotemark}
\makeatother

\newcommand{\mathset}[1]{\mathbb{#1}}

\newcommand{\vb}{\bvec{b}}

\newcommand{\vd}{\bvec{d}}

\newcommand{\vf}{\bvec{f}}

\newcommand{\vh}{\bvec{h}}

\renewcommand{\vr}{\bvec{r}}

\newcommand{\vw}{\bvec{w}}
\newcommand{\vx}{\bvec{x}}
\newcommand{\vy}{\bvec{y}}

\newcommand{\vR}{\bvec{R}}

\newcommand{\vX}{\bvec{X}}
\newcommand{\vY}{\bvec{Y}}

\newcommand{\abs}[1]{\left\lvert#1\right\rvert}

\newcommand{\ud}{\,\mathrm{d}}

\renewcommand{\EE}{\mathset{E}}

\newcommand{\RR}{\mathset{R}}

\theoremstyle{plain}

\theoremstyle{plain}

\theoremstyle{plain}
\newtheorem*{lem*}{\protect\lemmaname}
\theoremstyle{plain}

\theoremstyle{plain}

\providecommand{\definitionname}{Definition}
\providecommand{\assumptionname}{Assumption}
\providecommand{\corollaryname}{Corollary}
\providecommand{\lemmaname}{Lemma}
\providecommand{\propositionname}{Proposition}
\providecommand{\remarkname}{Remark}
\providecommand{\theoremname}{Theorem}

\usetikzlibrary{quantikz}

\usetikzlibrary{positioning, calc, fit}
\tikzset{%
  highlight/.style={rectangle,rounded corners,fill=blue!15,draw,fill opacity=0.3,thick,inner sep=0pt}
}

\usepackage[version=3]{mhchem}

\newcommand{\ee}{\mathrm{ee}}
\newcommand{\ei}{\mathrm{ei}}
\newcommand{\II}{\mathrm{II}}

\begin{document}

\title{Explicitly antisymmetrized neural network layers \\for variational Monte Carlo simulation}
\author{Jeffmin Lin}
\affiliation{Department of Mathematics, University of
California, Berkeley, CA 94720, USA}
\author{Gil Goldshlager}
\affiliation{Department of Mathematics, University of
California, Berkeley, CA 94720, USA}
\author{Lin Lin}
\email{linlin@math.berkeley.edu}
\affiliation{Department of Mathematics, University of
California, Berkeley, CA 94720, USA}
\affiliation{Computational Research Division, Lawrence Berkeley National Laboratory,
Berkeley, CA 94720, USA}
\date{\today}
\begin{abstract}
The combination of neural networks and quantum Monte Carlo methods has arisen as a promising path forward for highly accurate electronic structure calculations.
Previous proposals have combined equivariant neural network layers with a final antisymmetric layer in order to satisfy the antisymmetry requirements of the electronic wavefunction. However, to date it is unclear if one can represent antisymmetric functions of physical interest, and it is difficult to precisely measure the expressiveness of the antisymmetric layer.
This work attempts to address this problem by introducing explicitly antisymmetrized universal neural network layers. 
This approach has a computational cost which increases factorially with respect to the system size, but we are nonetheless able to apply it to small systems to better understand how the structure of the antisymmetric layer affects its performance.
We first introduce a generic antisymmetric (GA) neural network layer, which we use to replace the entire antisymmetric layer of the highly accurate ansatz known as the FermiNet.
We demonstrate that the resulting FermiNet-GA architecture can yield effectively the exact ground state energy for small atoms and molecules. We then consider a factorized antisymmetric (FA) layer which more directly generalizes the FermiNet by replacing the products of determinants with products of antisymmetrized neural networks. We find, interestingly, that the resulting FermiNet-FA architecture does not outperform the FermiNet. This strongly suggests that the sum of products of antisymmetries is a key limiting aspect of the FermiNet architecture. 
To explore this further, we investigate a slight modification of the FermiNet, called the full determinant mode, which replaces each product of determinants with a single combined determinant. We find that the full single-determinant FermiNet closes a large part of the gap between the standard single-determinant FermiNet and FermiNet-GA on small atomic and molecular problems. 
Surprisingly, on the nitrogen molecule at a dissociating bond length of 4.0 Bohr, the full single-determinant FermiNet can significantly outperform the largest standard FermiNet calculation with 64 determinants, yielding an energy within $0.4$ kcal/mol of the best available computational benchmark.
\end{abstract}
\maketitle

\section{Introduction}
A fundamental challenge in modeling the behavior of electrons in the many-body Schr\"odinger equation is that the electronic wavefunction must be antisymmetric with respect to particle exchange. When the number of electrons grows, effective parametrization of the space of such wavefunctions becomes difficult. Deep learning techniques have recently impacted \textit{ab initio} quantum chemistry by providing a new approach to the problem of tractable parameterization of high dimensional function spaces in quantum many-body problems. Over the past few years, a growing number of works \cite{Carleo2017, Nomura2017, Choo2018, Nagy2019, Luo2019, Han2019, Yang2020, Hermann2020, Pfau2020, Choo2020, StokesRobledo2020} have demonstrated the use of neural networks in wavefunction approximation, with an increasing amount of importance placed on building symmetry constraints into models. In particular, several works \cite{Luo2019, Han2019, Hermann2020, Pfau2020, StokesRobledo2020} have recently applied neural networks to model antisymmetric wavefunctions.

The simplest ansatz for representing antisymmetric electronic wavefunctions is known as a Slater determinant, which is an antisymmetrized product of single particle orbitals. The optimization of this ansatz is the core of the Hartree-Fock (HF) method~\cite{SzaboOstlund1989}. Conventionally, the representation power of the Slater determinant has been improved by including multiplicative Jastrow factors and transforming the particle coordinates via a so-called backflow transformation \cite{Feynman1956, Tocchio2008}, resulting in the Slater--Jastrow--backflow ansatz. While the Hartree Fock problem can be solved efficiently for a wide range of systems of interest using matrix diagonalization methods, the Slater--Jastrow and the Slater--Jastrow--backflow ansatzes are significantly more complicated and can in practice only be optimized using the quantum Monte Carlo (QMC) method known as variational Monte Carlo (VMC)~\cite{Foulkes2001,GubernatisKawashimaWerner2016,Toulouse2016,Becca2017}.
According to the variational principle, the energy obtained by any admissible wavefunction ansatz is lower bounded by the exact ground state energy.
For strongly correlated quantum systems, and even weakly correlated quantum
systems when high accuracy is required, a linear combination of either a large number
of Slater determinants (called the configuration interaction method (CI)) or a number of Slater--Jastrow--backflow ansatzes is needed to
yield a sufficiently low, and therefore accurate, energy estimate.

Recently, these considerations have led to an active interest in leveraging neural networks to improve the construction of the backflow \cite{Luo2019, Hermann2020, Pfau2020}, the antisymmetry \cite{Han2019, Pfau2020}, and the Jastrow factor \cite{Hermann2020} of these ansatzes. PauliNet \cite{Hermann2020} uses relatively small permutation equivariant neural networks for the backflow and invariant neural networks for the Jastrow factor. The backflow transformation in PauliNet is applied multiplicatively to the Hartree-Fock orbitals before the determinant layer is applied.
The FermiNet work \cite{Pfau2020, Spencer2020}, revealed around the same time as PauliNet, uses a more sophisticated equivariant backflow transformation with many more parameters. Another interesting and surprising feature of the FermiNet is that it eschews the Jastrow factor entirely. For a given system, the FermiNet often achieves lower energies than PauliNet~\cite{Spencer2020}. 

While there has been some progress~\cite{HanLiLinEtAl2019,SannaiTakaiCordonnier2019,KerivenPeyre2019,Hutter2020,BachmayrDussonOrtner2021} in analyzing the expressiveness of the permutation equivariant mappings used in the backflow construction~\cite{ZaheerKotturRavanbakhshEtAl2017}, the understanding of the effectiveness of the antisymmetric neural network layers remains limited~\cite{HanLiLinEtAl2019,Hutter2020,KesslerCalcavecchiaKuehne2021}.
Interestingly, Refs.~\cite{Luo2019,Pfau2020,Hutter2020} propose that a single FermiNet determinant could in theory achieve a universal representation of antisymmetric functions. 
However, these constructions are based on either a sorting process~\cite{Luo2019,Pfau2020} or an equivariant mapping that essentially encodes the entire wavefunction~\cite{Hutter2020}. Both constructions yield discontinuous feature mappings when the ambient space dimension is larger than $1$ or the number of particles is greater than $2$, as opposed to the continuous neural network layers used in all works in the literature so far.
Furthermore, in practice, the success of both PauliNet and FermiNet depends crucially on the quality of the permutation equivariant backflow. Therefore when the VMC energy is higher than the exact ground state energy, it is difficult to pin down the source of the error. 

To address this issue, we consider wavefunction ansatzes which replace parts of the antisymmetric layer of FermiNet with explicitly antisymmetrized universal neural networks.
The obvious drawback of this approach is that its computational cost increases factorially with respect to the number of electrons, and it can therefore only be applied to very small atoms and molecules. However, the use of explicit antisymmetrization can still allow us to better understand how the structure of the antisymmetric layer affects the overall performance.

We first consider a generic antisymmetric (GA) layer, which replaces the entire sum of products of determinants structure in FermiNet with an explicitly antisymmetrized feed forward neural network.
When combined with the FermiNet backflow, the resulting FermiNet-GA architecture can achieve a universal representation of antisymmetric functions by construction.
We also find that the FermiNet-GA structure is empirically a highly expressive ansatz.
For all systems studied, the error of the correlation energy is less than $1\%$, and is well below chemical accuracy ($1$ kcal/mol $\approx 1.6\times 10^{-3}$ a.u.). 
On the other hand, at least from a practical perspective, we find that the so-called single-determinant FermiNet, which is in fact computed as a product of two determinants, is not expressive enough to represent electronic wavefunctions of interest. 

To investigate further, we replace the product of determinants in the single-determinant FermiNet with a product of explicitly antisymmetrized feed forward neural networks, yielding the factorized antisymmetric neural network layer of rank $1$ (FA-$1$). We find that FA-$1$ is not able to outperform single-determinant FermiNet, which suggests that the ineffectiveness of the single-determinant FermiNet is closely related to its product structure. To explore this further, we define a factorized antisymmetric neural network layer of rank $K$ (FA-$K$), which generalizes the structure of $K$-determinant FermiNet. We find that FermiNet-FA-$K$ does not outperform $K$-determinant FermiNet, which indicates that the sum-of-products structure is a key limiting feature of both architectures.

These results suggest that removing this sum-of-products structure may be a promising avenue towards developing an efficient antisymmetric layer that is more expressive than the original FermiNet architecture. We thus study a variant of the FermiNet called full determinant mode which replaces the products of determinants used in the FermiNet with single combined determinants. The full determinant construction is implemented in the JAX branch of the FermiNet repository \cite{Spencer2020}, and is also mentioned  in passing in the original FermiNet paper \cite{Pfau2020}, but to our knowledge its performance has not been reported in the literature. We specifically investigate the full single-determinant FermiNet, which replaces the product of two determinants used in single-determinant FermiNet with a single combined determinant. 
Our numerical results show that full single-determinant FermiNet can close a large part of the gap between standard single-determinant FermiNet and the true ground state energy on small atomic and molecular problems.
We further evaluate the performance of full single-determinant FermiNet on the nitrogen molecule at a dissociating bond length of 4.0 Bohr, a challenging strongly correlated system where the standard FermiNet architecture is not able to yield accurate results even with 64 determinants.
To our great surprise, we find that the full single-determinant FermiNet can outperform the standard 64-determinant FermiNet on this system, and the error of the energy can be as small as $0.4$ kcal/mol compared to the best available computational benchmark.

\section{Preliminaries}

\subsection{Many-body electron problem}
We consider isolated quantum chemical systems in $\RR^3$ and use atomic units (a.u.) throughout the paper. Let $\{\vr_i\}_{i=1}^N$ represent electron positions, $\{\vR_I^{(a)}\}_{i=1}^M$ nuclei positions, and $Z_I$ the $I$th nuclear charge. Under the Born-Oppenheimer approximation, the many body Hamiltonian with $M$ nuclei and $N$ electrons is
\begin{equation}
H =  \hat{T} + \hat{V}_{\ei} + \hat{V}_{\ee} + E_{\II},
\label{eqn:manybody_ham}
\end{equation}
where the Hamiltonian is partitioned into the kinetic, electron-ion potential, electron-electron, and ion-ion interaction, respectively:
\begin{equation}
\hat{T} = \sum_{i=1}^N -\frac{1}{2} \Delta_{\vr_i}, \quad \hat{V}_{\ei} =
-\sum_{i=1}^N \sum_{I=1}^M \frac{Z_I}{\abs{\vr_i - \vR^{(a)}_I}}, \quad \hat{V}_{\ee} = \sum_{i < j}^N \frac{1}{\abs{\vr_i -
\vr_j}}, \quad E_{\II} = \sum_{I < J}^M \frac{Z_I Z_J}{\abs{\vR^{(a)}_I - \vR^{(a)}_J}}.
\end{equation}
For a given atomic configuration $\{\vR^{(a)}_{I}\}$, the ion-ion interaction simply adds a constant shift to the Hamiltonian.

Let $\vX=(\vx_1,\ldots,\vx_N)$ be the collection of  spatial and spin indices of $N$ electrons, where $\vx_i=(\vr_i,\sigma_i)$, and assume the $N$ electrons can be partitioned into $N_{\uparrow}$ spin-up electrons and $N_{\downarrow}$ spin-down electrons.
We denote by $\vR=(\vr_1,\ldots,\vr_N)$ the collection of spatial coordinates of the electrons. We also denote by $\vR^{(a)}=(\vR^{(a)}_1,\ldots,\vR^{(a)}_M)$ the collection of atomic positions. Then the electron many-body wavefunction $\Psi(\vX) \equiv \Psi(\vx_1,\ldots,\vx_N)$
is required to be antisymmetric with respect to the natural action of the symmetric group $S_N$ on the electron configuration $\vX$ (i.e., the permutation of the particle indices):
\begin{equation}
\Psi(\pi(\vX)) \equiv \Psi(\vx_{\pi(1)},\ldots,\vx_{\pi(N)}) = (-1)^\pi \Psi(\vX),
\end{equation}
where $(-1)^\pi$ is the sign or parity of the permutation $\pi\in S_N$.

Note that the Hamiltonian in \cref{eqn:manybody_ham} does not explicitly depend on the spin. As a result, the antisymmetry constraint over the spatial-spin electron configuration can be rewritten using a spin-independent wavefunction $\Psi(\vR)$ (abusing notation to reuse $\Psi$), which reduces the number of degrees of freedom in the wavefunction and improves the efficiency in quantum Monte Carlo calculations~\cite{Foulkes2001}.

To see this, let $\hat{O}$ be any totally symmetric spin-independent operator (e.g. electron density, energy), and assume we are interested in expectation values of $\hat{O}$ with respect to the many body wavefunction $\Psi(\vX)$:
\begin{equation}
\braket{\hat{O}}=\frac{\sum_{\sigma} \int \Psi^{*}(\vX) \hat{O}(\vR) \Psi(\vX) \ud \vR}{\sum_{\sigma} \int \Psi^{*}(\vX) \Psi(\vX) \ud \vR}.
\label{eqn:Oaverage}
\end{equation}

For each spin-configuration $\sigma=(\sigma_1,\ldots,\sigma_N)$, we can always permute the particle index $1,\ldots,N$ so that the spin-up indices appear in front of the spin-down indices. The integrals in the numerator and denominator of \cref{eqn:Oaverage} are independent of such a permutation operation. Therefore we may define the spatial wavefunction
\begin{equation}
\Psi(\vR)=\Psi((\vr_1,\uparrow),\ldots,(\vr_{N_{\uparrow}},\uparrow), 
(\vr_{N_{\uparrow}+1},\downarrow),\ldots,(\vr_{N_{\uparrow}+N_{\downarrow}},\downarrow)).
\label{eqn:psi_spatial}
\end{equation}
By renaming the integration variables,
\begin{displaymath}
\begin{aligned}
\braket{\hat{O}}=&\frac{\sum_{\sigma} \int \Psi^*(\vR) \hat{O}(\vR) \Psi(\vR) \ud \vR}{\sum_{\sigma} \int \Psi^*(\vR) \Psi(\vR) \ud \vR}\\
=&\frac{ \int \Psi^*(\vR) \hat{O}(\vR) \Psi(\vR) \ud \vR}{ \int \Psi^*(\vR) \Psi(\vR) \ud \vR}.
\end{aligned}
\end{displaymath}
We further introduce the notation $\vR^{\uparrow}\equiv(\vr_1^{\uparrow},\ldots,\vr_{N_{\uparrow}}^{\uparrow}) \equiv (\vr_1,\ldots,\vr_{N_\uparrow})$ and $\vR^{\downarrow}\equiv(\vr_1^{\downarrow},\ldots,\vr_{N_{\downarrow}}^{\downarrow}) \equiv (\vr_{N_\uparrow + 1},\ldots,\vr_{N_\uparrow + N_\downarrow})$. Then the constraint that $\Psi(\vX)$ is antisymmetric with respect to the action of $S_N$ is equivalent to the requirement that
\begin{equation}
\Psi(\vR):=\Psi(\vR^{\uparrow},\vR^{\downarrow})
\label{eqn:phi_R}
\end{equation}
is an antisymmetric function with respect to $\vR^{\uparrow}$ and $\vR^{\downarrow}$ separately, but not necessarily antisymmetric across the spin-up and spin-down indices. We can thus work with $\Psi(\vR)$ instead of $\Psi(\vX)$ as long as we enforce this symmetry. If needed, the original antisymmetric function $\Psi(\vX)$ can be recovered by antisymmetrizing the following spin-dependent function:
\begin{equation}
\Psi(\vR) \delta_{\sigma_{1}, \uparrow} \ldots \delta_{\sigma_{N_{\uparrow}}, \uparrow} \delta_{\sigma_{N_{\uparrow}+1}, \downarrow} \cdots \delta_{\sigma_{N}, \downarrow}.
\end{equation}

\subsection{Variational Monte Carlo}
In the variational Monte Carlo (VMC) method, the goal is to find the variational minimizer $\Psi$ of the energy functional
\begin{equation}
E[\Psi] = \frac{\int \Psi^*(\vR) H\Psi(\vR)\,d\vR}{\int \Psi^*(\vR)\Psi(\vR)\,d\vR}.
\label{eqn:energy_functional}
\end{equation}

Given a wavefunction $\Psi$, which may not be normalized, we can estimate the high-dimensional integral in \cref{eqn:energy_functional} via Markov-Chain Monte Carlo sampling. We first define the probability distribution associated with $\Psi(\vR)$ as $p(\vR) = \left|\Psi(\vR)\right|^2 / \int \left|\Psi(\vR)\right|^2\,d\vR$. We then write the energy functional as a simple expectation over $p(\vR)$:
\begin{equation}
    \frac{\int \Psi^*(\vR) H\Psi(\vR)\,d\vR}{\int \Psi^*(\vR)\Psi(\vR)\,d\vR}
    = \int_{\RR^{Nd}} \frac{H\Psi(\vR)}{\Psi(\vR)}\,p(\vR)\,d\vR
    = \int_{\RR^{Nd}} E_L(\vR)p(\vR)\,d\vR,
\end{equation} 
where we have defined the so-called \textit{local energy} associated with $\Psi$ as 
\begin{equation}
    E_{L}(\vR) = \frac{H\Psi(\vR)}{\Psi(\vR)}.
\end{equation}

The general strategy for solving the molecular many-body problem \eqref{eqn:manybody_ham} is then to parametrize a family of real wavefunctions $\Psi_\theta\in L^2$ on some parameter domain (usually $\RR^m$, where $m$ is the number of parameters) and combine the parameterization with Monte Carlo estimation to solve the resulting approximate eigenproblem. This is done by drawing a set of samples $\xi_\theta$ from $p_\theta(\vR)$ using Markov-Chain Monte Carlo and estimating the loss as 
\begin{equation}
\mathcal{L}(\theta) = E[\Psi_\theta] \approx \frac{1}{|\xi_{\theta}|}\sum_{\vR\in \xi_{\theta}} E_L(\vR;\theta) = \widetilde{\mathcal{L}}(\theta).
\label{eqn:loss_def}
\end{equation}
In statistical machine learning parlance, $\widetilde{\mathcal{L}}$ is known as the empirical risk, and $\mathcal{L}$ is known as the true or population risk.

\section{Architectures}
\pgfdeclarelayer{bg}    
\pgfsetlayers{bg,main}  

\newcommand{\vecheight}{0.7cm}
\newcommand{\vecwidth}{1.1}
\newcommand{\aspace}{0.15}
\newcommand{\spinheight}{0.75}
\newcommand{\vdotoffset}{0.105}

\newcommand{\plusSign}{\large $+$}
\newcommand{\timesSign}{\large $\times$}

\newcommand{\scalarColor}{green!20}
\newcommand{\vectorColor}{yellow!40}
\newcommand{\signColor}{purple!25}
\newcommand{\opColor}{blue!25}

\newcommand{\yvec}[7]{
\begin{small} 
\node[draw, minimum width = \vecwidth cm, minimum height = 4.5*\vecheight, fill=\vectorColor] (yvec#7) at (#5 + \vecwidth/2, #6 + \vecheight * 2.25) {};
\node at (#5 + \vecwidth/2, #6 + \vecheight*4) {$\vy^#1_{#2}$};
\draw (#5, #6 + \vecheight * 3.5) -- (#5 + \vecwidth, #6 + \vecheight*3.5);
\node at (#5 + \vecwidth/2, #6 + \vecheight*3) {$\vy^#1_{#3}$};
\draw (#5, #6 + \vecheight*2.5) -- (#5 + \vecwidth, #6 + \vecheight*2.5);
\node at (#5 + \vecwidth/2, #6 + \vecheight*1.75 + \vdotoffset cm) {\large $\textbf{\vdots}$};
\draw (#5, #6 + \vecheight) -- (#5 + \vecwidth, #6 + \vecheight);
\node at (#5 + \vecwidth/2, #6 + 0.5*\vecheight) {$\vy^#1_{#4}$};
\end{small}
}

\newcommand{\yvecperm}[4]{
\yvec{#2}{1}{2}{N_#2}{0}{#3}{#4};
\yvec{#2}{#1(1)}{#1(2)}{#1(N_#2)}{1.75}{#3}{#4perm};
\draw[->, >=latex] (yvec#4) -- (yvec#4perm) node[midway,above] {\large $#1$};
}

\newcommand{\Fperm}[1]{
\yvecperm{\pi}{\sigma}{0}{}

\node[draw, right = 1.5 of yvecperm, fill=\scalarColor] (Xi) {\large $\Xi_{\mathrm{FFNN}}^\sigma(#1(\vY^\sigma))$};
\draw[->, >=latex] (yvecperm) -- (Xi) node[midway,above] {FFNN};

\node[draw, below = .5 of Xi,fill=\signColor] (Sign) {\large $\mathrm{sgn}(#1)$};
\node[draw, circle, right = .75 of Xi, fill=\opColor] (Times) {\timesSign};

\draw[->, >=latex] (Sign) -- (Times);
\draw[->, >=latex] (Xi) -- (Times);

\node[draw, fill=\scalarColor, right = 0.75 of Times] (Fperm) {\Large $F^\sigma_#1(\vY^\sigma)$};
\draw[->, >=latex] (Times) -- (Fperm);
\node[draw, dashed, fit={(Fperm) (yvec)}, inner sep = 0.25cm] (FPermBox) {};
}

\newcommand{\FAspin}[4]{
\node[draw, dashed, fill=\vectorColor] (Y#4) at (#3,#2)  {\Large $\vY^#1$} ;

\node[draw, dashed,  fill=\scalarColor, minimum width = 2.5cm, , minimum height = 0.7cm] (F1#4) at (2 + #3, #2 + 2*\spinheight)  {$F^#1_{\pi_1}(\vY^#1)$};
\node[draw, fill=\scalarColor, minimum width = 2.5cm, , minimum height = 0.7cm] (F2#4) at (2 + #3, #2 + 1*\spinheight)  {$F^#1_{\pi_2}(\vY^#1)$};
\node at (2 + #3, #2 - 0.5*\spinheight + \vdotoffset)  {\large $\textbf{\vdots}$};
\node[draw, fill=\scalarColor, minimum width = 2.5cm, , minimum height = 0.7cm] (FN!#4) at (2 + #3, #2 -2*\spinheight)  {$F^#1_{\pi_{N_#1!}}(\vY^#1)$};

\draw[dashed,  ->, >=latex] (Y#4) -- (F1#4.west);

\draw[->, >=latex] (Y#4) -- (F2#4.west);
\draw[->, >=latex] (Y#4) -- (FN!#4.west);

\node[draw, circle, fill=\opColor, right = 3.25 of Y#4] (plus#4) {\plusSign};

\draw[->, >=latex] (F1#4.east) -- (plus#4);
\draw[->, >=latex] (F2#4.east) -- (plus#4);
\draw[->, >=latex] (FN!#4.east) -- (plus#4);

\node[draw, right=0.75 of plus#4, fill=\scalarColor] (Psi#4) { \large $\mathcal{A}_#1[\Xi_{\mathrm{FFNN}}^#1](\vY^#1)$};

\draw[->, >=latex] (plus#4) -- (Psi#4);
}

\newcommand{\FA}[2]{
\FAspin{\uparrow}{#1 + 2}{#2}{up}
\FAspin{\downarrow}{#1 - 2}{#2}{down}

\node[draw, circle, fill=\opColor] (Times) at ($(Psiup) !0.5! (Psidown) + (2,0)$) {\timesSign};

\draw[->, >=latex] (Psiup.south east) -- (Times);
 
\begin{pgfonlayer}{bg}    
\draw[dotted] (Yup.north west) -- (FPermBox.south west);
\draw[dotted] (F1up.north east) -- (FPermBox.south east);
\draw[dotted] (Ydown.north west) -- (FPermBox.south west);
\draw[dotted] (F1down.north east) -- (FPermBox.south east);
\end{pgfonlayer}

\draw[->, >=latex] (Psidown.north east) -- (Times);

\node[right = 1 of Times] (Psi) {\huge $\Psi$};

\draw[->, >=latex] (Times) -- (Psi);
}

\newcommand{\FGA}{
\yvecperm{\pi}{\uparrow}{3.5cm}{pi}
\yvecperm{\rho}{\downarrow}{0}{rho}

\draw (yvecpiperm) -- +(.75, 0);
\draw (yvecrhoperm) -- +(.75, 0);
\draw ($(yvecrhoperm) + (.75, 0)$) --  ($(yvecpiperm)+(.75, 0)$);

\node[draw, fill=\scalarColor] (Xi) at ($(yvecrhoperm) !0.25! (yvecpiperm) + (3.75,0)$) { $\Xi_{\mathrm{FFNN}}(\pi(\vY^\uparrow),\rho(\vY^\downarrow))$};

\draw[->, >=latex] ($(yvecrhoperm) !0.25! (yvecpiperm) + (.75,0)$) -- (Xi) node[midway,above] {FFNN};

\node[draw, circle, above = 0.5 of Xi, fill=\opColor] (Times) {\timesSign};
\node[left  =0.5 of Times, draw, fill=\signColor] (SignPi) { $\mathrm{sgn}(\pi)$};
\node[right =0.5 of Times, draw, fill=\signColor] (SignRho) { $\mathrm{sgn}(\rho)$};

\draw[->, >=latex] (Xi) -- (Times);
\draw[->, >=latex] (SignPi) -- (Times);
\draw[->, >=latex] (SignRho) -- (Times);

\node[draw, fill=\scalarColor, above = 0.5 of Times] (Fperm) {\Large $F_{\pi, \rho}(\vY)$};
\draw[->, >=latex] (Times) -- (Fperm);
\node[draw, dashed, fit={(Fperm) (yvecpi) (yvecrho) (SignRho)}, inner sep = 0.25cm] (FPermBox) {};
}

\newcommand{\GAspin}[2]{
\node[draw, fill=\vectorColor, dashed] (Y) at (#2,#1)  {\Large $\vY$} ;

\node[draw, dashed, fill=\scalarColor, minimum width = 2.5cm, , minimum height = 0.7cm] (F1) at (#2 + 1.875, #1 + 3*\spinheight)  {$F_{\pi_1, \rho_1}(\vY)$};
\node[draw, fill=\scalarColor, minimum width = 2.5cm, , minimum height = 0.7cm] (F2) at (#2 + 1.875, #1 + 2*\spinheight)  {$F_{\pi_1, \rho_2}(\vY)$};

\node at (#2 + 1.875, #1 + 1*\spinheight + \vdotoffset)  {\large $\textbf{\vdots}$};

\node[draw, fill=\scalarColor, minimum width = 2.5cm, , minimum height = 0.7cm] (F3) at (#2 + 1.875, #1)  {$F_{\pi_2, \rho_1}(\vY)$};
\node[draw, fill=\scalarColor, minimum width = 2.5cm, , minimum height = 0.7cm] (F4) at (#2 + 1.875, #1 -1*\spinheight)  {$F_{\pi_2, \rho_2}(\vY)$};

\node at (#2 + 1.875, #1 - 2*\spinheight + \vdotoffset)  {\large $\textbf{\vdots}$};

\node[draw, fill=\scalarColor, minimum width = 2.5cm, minimum height = 0.7cm] (FN!) at (#2 + 1.875, #1 - 3*\spinheight)  {$F_{\pi_{N_\uparrow!},\rho_{N_\downarrow!}}(\vY)$};

\draw[->, >=latex, dashed, line width = 0.8] (Y) -- (F1.west);
\draw[->, >=latex] (Y) -- (F2.west);
\draw[->, >=latex] (Y) -- (F3.west);
\draw[->, >=latex] (Y) -- (F4.west);
\draw[->, >=latex] (Y) -- (FN!.west);

\node[draw, circle, fill=\opColor, right = 3 of Y] (plus) {\plusSign};

\draw[->, >=latex] (F1.east) -- (plus);
\draw[->, >=latex] (F2.east) -- (plus);
\draw[->, >=latex] (F3.east) -- (plus);
\draw[->, >=latex] (F4.east) -- (plus);
\draw[->, >=latex] (FN!.east) -- (plus);

\node[right =0.25 of plus] (Psi) {\huge $\Psi$};

\draw[->, >=latex] (plus) -- (Psi);
}

\newcommand{\FAdiagram}{
\begin{tikzpicture}
\Fperm{\pi}
\FA{-4.5}{.75}
\end{tikzpicture}
}

\newcommand{\GAdiagram}{
\begin{tikzpicture}
\FGA
\GAspin{3.325}{9.375}

\begin{pgfonlayer}{bg} 
\draw[dotted] (Y.south west) -- (FPermBox.south east);
\draw[dotted] (F1.north east) -- (FPermBox.north east);
\end{pgfonlayer}
\end{tikzpicture}
}
\label{section:architecture}
\newcommand{\orbital}[2]{\varphi^{#1\sigma}_{#2}}
\newcommand{\orbitalfn}[3]{\orbital{#1}{#2}(\vy^\sigma_{#3})}
\newcommand{\orbmat}[2][\sigma]{\Phi^{#2#1}}
\newcommand{\orbmatfull}[2][\sigma]{\Phi_{\mathrm{full}}^{#2#1}}
\newcommand{\orbmatfn}[2][\sigma]{\orbmat[#1]{#2}(\vY^{#1})}
\newcommand{\orbmatfnfull}[2][\sigma]{\orbmatfull[#1]{#2}(\vY^{#1})}

In this section, we describe the architectures for the neural network-based ansatzes that we explore. We first describe the FermiNet architecture, which consists of a (generalized) backflow layer that produces permutation equivariant features, followed by an antisymmetric layer to compute the wavefunction amplitude. We then describe the other architectures that we explore, which use different antisymmetric layers but all share the same generalized backflow layer used in the FermiNet.

The overall structure of the FermiNet may be described as a composition of a general $S_{N_{\uparrow}}\times S_{N_{\downarrow}}$ equivariant feature map (see the definition below) $\vY(\vR) \equiv (\vY^\uparrow(\vR), \vY^\downarrow(\vR))$, with an antisymmetric layer constructed as a sum of products of determinants of orbital matrices $\Phi^{k\sigma}$:
\begin{equation}
\Psi(\vR) = \sum_{k=1}^K \det \Phi^{k\uparrow}(\vY^{\uparrow}(\vR)) \det \Phi^{k\downarrow}(\vY^{\downarrow}(\vR)).
\label{eqn:ferminet_overall}
\end{equation}
We describe the equivariant feature map and the orbital-determinant layer separately below.

\subsection{Permutation equivariant features in FermiNet}
The FermiNet feature map $\vY$ is a map from particle positions $\vR\in\RR^{N\times d}$ to generalized coordinates $\vY\in\RR^{N\times d'}$.
While $d=3$ is the dimension of the physical space, $d'$ is the number of features and can be chosen arbitrarily.
One particularly important symmetry of interest is given by the tensor product of the canonical permutation representations of $S_{N_\uparrow}$ and $S_{N_\downarrow}$, defined by the action
\begin{equation}
(\pi\otimes\rho)(\vR) = (\pi\otimes\rho)(\vR^{\uparrow},\vR^{\downarrow}) = (\pi(\vR^{\uparrow}),\rho(\vR^{\downarrow})) = (\vr_{\pi(1)}^\uparrow,\ldots,\vr_{\pi(N_{\uparrow})}^\uparrow,\vr_{\rho(1)}^\downarrow,\ldots,\vr_{\rho(N_{\downarrow})}^{\downarrow}).
\end{equation}
The FermiNet map $\vY$ is \textit{equivariant} with respect to this action, i.e.
\begin{equation}
\vY((\pi\otimes\rho)(\vR)) = (\pi\otimes\rho)(\vY(\vR)).
\end{equation}

To ensure that $\vY$ is equivariant with respect to elements of $S_{N_{\uparrow}}\times S_{N_{\downarrow}}$ while incorporating information about the two-electron distances, the basic layer involves two streams: a ``one-electron stream'' that starts with the electron positions $\vR = (\vr_1,\ldots\vr_N)$, and a ``two-electron stream'' that starts with the electron-electron displacements $\vr_{ij} \equiv \vr_i - \vr_j$. The streams are averaged over the electrons, concatenated onto the one-electron stream, and a dense layer followed by a nonlinear activation function such as $\tanh$ is applied. Residual connections~\cite{HeZhangRenEtAl2016} are also used between layers of the same shape for both streams.

More precisely, if $\vh_i^{l\alpha}$ and $\vh_{ij}^{l\alpha\beta}$ are the outputs of the one- and two- electron streams at layer $l$ with spins $\alpha,\beta\in\{\uparrow,\downarrow\}$, then the concatenated vector for index $i$ and spin $\alpha$ is
\begin{equation}
\vf_i^{l\alpha} = \left(\vh_i^{l\alpha}, \frac{1}{N_{\uparrow}}\sum_{j=1}^{N_{\uparrow}}\vh_j^{l\uparrow}, \frac{1}{N_{\downarrow}}\sum_{j=1}^{N_{\downarrow}}\vh_j^{l\downarrow},\frac{1}{N_{\uparrow}}\sum_{j=1}^{N_{\uparrow}} \vh_{ij}^{l\alpha\uparrow}, \frac{1}{N_{\downarrow}}\sum_{j=1}^{N_{\downarrow}} \vh_{ij}^{l\alpha\downarrow}\right)
\label{eqn:one_elec_stream_concat}
\end{equation}
and the output of layer $l+1$ is given by the two streams
\begin{equation}
\begin{split}
\vh_i^{(l+1)\alpha} &= \tanh\left(V^l \vf_i^{l\alpha} + \vb^{l}\right) + \vh_i^{l\alpha},\\
\vh_{ij}^{(l+1)\alpha\beta} &= \tanh\left(W^l \vh_{ij}^{l\alpha\beta} + \vb^{l}\right) + \vh_{ij}^{l\alpha\beta}.
\end{split}
\label{eqn:ferminet_stream_def}
\end{equation}
As an initial pre-processing step, the electron positions $\vR$ are converted to ``atomic coordinates'' as
\begin{align}
\vh_i^{0\alpha} &= \left(\vr^{\alpha}_i - \vR_1^{(a)}, \abs{\vr^{\alpha}_i - \vR_1^{(a)}}, \ldots, \vr^{\alpha}_i - \vR_M^{(a)}, \abs{\vr^{\alpha}_i - \vR_M^{(a)}}\right),\\
\vh_{ij}^{0\alpha\beta} &= \left(\vr^{\alpha}_i - \vr^{\beta}_j, \abs{\vr^{\alpha}_i - \vr^{\beta}_j}\right),
\label{eqn:atomic_coords}
\end{align}
which are invariant with respect to a simultaneous translation of the entire system. The explicit dependence on the absolute values enables the network to efficiently represent the derivative discontinuity due to the electron-nuclei cusp and electron-electron cusp, respectively~\cite{Kato1957,Pfau2020}.

Each map $(\vh^{l\uparrow},\vh^{l\downarrow}) \mapsto (\vh^{(l+1)\uparrow},\vh^{(l+1)\downarrow})$ is $S_{N_\uparrow}\times S_{N_\downarrow}$ equivariant due to the averaging procedure~\cite{ZaheerKotturRavanbakhshEtAl2017} in \cref{eqn:one_elec_stream_concat}, and the map $\vR\mapsto \vh^0$ is parallel in the particle index $i$ and thus equivariant as well.
Therefore the map $\vR\mapsto \vY \equiv (\vY^\uparrow, \vY^\downarrow) \equiv (\vh^{L\uparrow}, \vh^{L\downarrow})$, where $L$ is the total number of one-electron layers, is also $S_{N_\uparrow}\times S_{N_\downarrow}$ equivariant. This is called a backflow map, which generalizes the original proposal of ``backflow'' by Feynman and Cohen \cite{Feynman1956}. 

Note that both $\vY^\uparrow$ and $\vY^\downarrow$ depend on the positions of all electrons. 
The index $\alpha$ in the notation $\vY^\alpha$ does not denote a dependence of $\vY$ only on the corresponding $\alpha$-spin inputs $\vR^\alpha$, but rather denotes the symmetry constraint of $\vY^\alpha$ as equivariant with respect to the action of $S_{N_\alpha}$ on $\vR^\alpha$ and invariant with respect to the action of $S_{N_\beta}$ on $\vR^{\beta}$. Hence in this paper, we shall refer to the index $\{\uparrow,\downarrow\}$ in the expression $(\vY^\uparrow,\vY^\downarrow)$ as the \textit{pseudospin} index. It has been shown constructively, though without explicit error bounds, that a simplified version of this construction~\cite{Hutter2020} can approximate all and only the equivariant continuous functions.

\subsection{Antisymmetric layer in FermiNet}

Once the equivariant feature maps $\vec{Y} = (\vec{Y}^\uparrow, \vec{Y}^\downarrow)$ are generated, they are then used to generate $K$ pairs of orbital matrices $(\orbmatfn[\uparrow]{k}, \orbmatfn[\downarrow]{k})$. These orbital matrices are constructed using a set of per-pseudospin single particle orbitals $\orbital{k}{i}$ ($\sigma\in\{\uparrow,\downarrow\}$), which are defined by applying a simple dense layer to the equivariant features and multiplying by an exponential envelope function:
\begin{equation}
\orbitalfn{k}{i}{j}=(\vw^{k\sigma}_{i}\cdot \vy^\sigma_j +\vb^{k\sigma}_{i}) \sum_{I} d^{k\sigma}_{i,I} \exp\left(-\abs{\mathbf{\Sigma}^{k\sigma}_{i,I}(\vr_j-\vR^{(a)}_I)}\right).
\label{eqn:phi_formulation}
\end{equation}
Here $\vw^{k\sigma}_{i}\in\RR^m$ and $b^{k\sigma}_{i}\in\RR$ are the weights and biases of the dense layer, where $m$ is the number of dimensions in the equivariant features $\vy^\sigma_i$, which is chosen to be constant for all $i,\sigma$. The exponential envelopes are parameterized by $d^{k\sigma}_{i,I}\in \RR$ and a matrix $\mathbf{\Sigma}^{k\sigma}_{i,I}\in\RR^{d\times d}$, which can also be set to a scaled identity matrix to simplify the ansatz. The parameters $\vw,\vb,\mathbf{\Sigma},\vd$ are trainable and dependent on the pseudospin index $\sigma$. The exponential term ensures that the wavefunction is normalizable and that the support of each orbital does not extend too far away into the vacuum. While the orbital functions $\orbital{k}{i}$ are applied only to single feature vectors $\vy_{j}$, the ansatz can express complex correlations because the feature vectors themselves depend on all of the particles in a complex way.

The orbital matrices follow naturally from these single particle orbitals as
\begin{equation}
\orbmatfn{k}=
\begin{pmatrix}
\orbitalfn{k}{1}{1} & \cdots & \orbitalfn{k}{1}{N_\sigma} \\
\vdots  & \ddots & \vdots \\
\orbitalfn{k}{N_\sigma}{1} & \cdots & \orbitalfn{k}{N_\sigma}{N_\sigma} \\
\end{pmatrix}
\label{eqn:orbital_matrix_one_spin}.
\end{equation}
Once these orbital matrices are constructed, FermiNet creates an antisymmetric wavefunction using a sum of products of determinants:

\begin{equation}
\Psi(\vR) = \sum_{k} \det \orbmat[\uparrow]{k} \det \orbmat[\downarrow]{k}
\end{equation}
These determinants are similar to Slater determinants except that their orbitals are general equivariant functions of all of the input particles rather than simple single particle orbitals. Such determinants are thus referred to as generalized Slater determinants. In fact, the original FermiNet architecture \cite{Pfau2020} used a weighted sum of products of generalized Slater determinants by adding in a set of trainable parameters $w_k$ and letting 
\begin{equation}
\Psi(\vR) = \sum_{k} w_k \det \orbmat[\uparrow]{k} \det \orbmat[\downarrow]{k}.
\end{equation}
However, these trainable weights were removed by the original authors in their follow-up work \cite{Spencer2020}, as they do not add extra expressiveness on top of the ability to tune the scale of the orbital matrices themselves. We follow the simplified construction.

\subsection{Generic antisymmetric neural network layer}

\begin{figure}
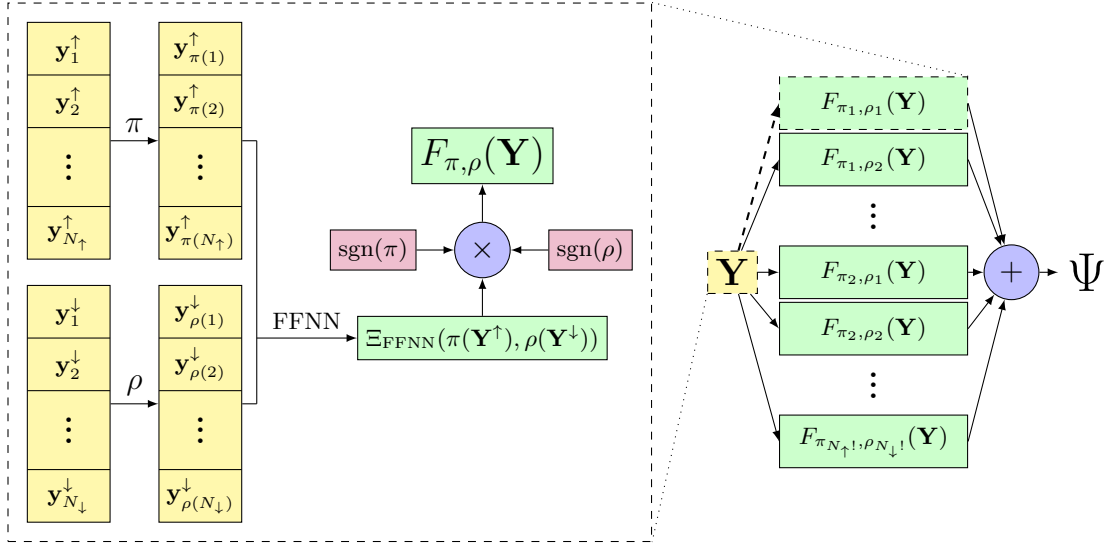

\GAdiagram
\caption{The architecture for the generic antisymmetric layer. Left: calculation of the wavefunction contribution from a single pair of permutations $\pi$ and $\rho$, denoted by $F_{\pi, \rho}(\vY)$. Right: combination of contributions for all permutations $\pi$ and $\rho$. This is implemented as a batch calculation over all pairs of permutations, but we show separate arrows for each pair of permutations to emphasize the factorial complexity of the operation.}
\label{diagram:GA}
\end{figure}

In order to assess the effectiveness of the antisymmetric layer of the FermiNet, we can replace the product of pseudospin determinant terms with a truly universal antisymmetric neural network layer. Let us first define the antisymmetrization operator on spin $\sigma$ as
\begin{equation}
\mathcal{A}_\sigma[f](\vR) \equiv \sum_{\pi_\sigma\in S_{N_\sigma}} (-1)^{\pi_\sigma}f(\pi_\sigma(\vR)),
\label{eqn:antisymmetrization_op}
\end{equation}
where $S_{N_\sigma}$ is the symmetric group on $\{1, \ldots, N_\sigma\}$. Following the Leibniz formula for the determinant, the antisymmetric structure for the standard single-determinant FermiNet can be written as
\begin{equation}
\begin{split}
\Psi(\vR) &=  \det \Phi^{\uparrow}(\vY^{\uparrow}(\vR))\cdot
\det \Phi^{\downarrow}(\vY^{\downarrow}(\vR)) \\
&= \prod_{\sigma\in\{\uparrow,\downarrow\}}\left(\sum_{\pi_{\sigma}\in S_{N_{\sigma}}}
(-1)^{\pi_{\sigma}}\varphi^\sigma_{1}(\vy^\sigma_{\pi_\sigma(1)})\cdots \varphi^\sigma_{N_\sigma}(\vy^\sigma_{\pi_\sigma(N_\sigma)})\right), \\
&\equiv \prod_{\sigma\in\{\uparrow,\downarrow\}} \mathcal{A}_\sigma[\Xi_{\mathrm{FermiNet}}^\sigma](\vY^\sigma),
\end{split}
\label{eqn:determinant_as_antisymmetrized_product}
\end{equation}
where we have defined the orbital product function $\Xi_{\mathrm{FermiNet}}^\sigma(\vY^\sigma) = \prod_{i=1}^{N_\sigma}\varphi_i^\sigma(\vy^\sigma_i, \theta_i^\sigma)$.

However, we can treat the interaction between pseudospins in a more universal way by considering just a single function $f(\vR^\uparrow,\vR^\downarrow)$ of all electron positions and antisymmetrizing $f$ with respect to only the subsets of the input indices that correspond to the two spins: 
\begin{equation}
\mathcal{A}_\uparrow\mathcal{A}_\downarrow[f](\vR^\uparrow, \vR^\downarrow) = \sum_{\pi_\uparrow\in S_{N_\uparrow}} \sum_{\pi_\downarrow\in S_{N_\downarrow}} (-1)^{\pi_\uparrow}(-1)^{\pi_\downarrow}f(\pi_{\uparrow}(\vR^\uparrow), \pi_{\downarrow}(\vR^\downarrow)).
\label{eqn:ga_op}
\end{equation}
The standard FermiNet can be written as the doubly antisymmetrized product of orbitals over all particles
\begin{equation}
\Psi_{\mathrm{FermiNet}}(\vR) = \mathcal{A}_\uparrow\mathcal{A}_\downarrow \left[\Xi_{\mathrm{FermiNet}}\right](\vY^\uparrow, \vY^\downarrow),
\end{equation}
where we define the product over all orbitals
\begin{equation}
\Xi_{\mathrm{FermiNet}}(\vY^\uparrow, \vY^\downarrow) = \varphi_1^\uparrow(\vy_1^\uparrow)\cdots\varphi^\uparrow_{N_\uparrow}(\vy^\uparrow_{N_\uparrow})\varphi_1^\downarrow(\vy^\downarrow_2)\cdots\varphi^\downarrow_{N_\downarrow}(\vy^\downarrow_{N_\downarrow}).
\end{equation}
If we replace this all-particle product with a single feed-forward neural network (FFNN) denoted by $\Xi_{\mathrm{FFNN}}(\vY^\uparrow, \vY^\downarrow)$, we arrive at an architecture that we refer to as the generic antisymmetric layer (GA):
\begin{equation}
\Psi_{\mathrm{GA}}(\vR) = \mathcal{A}_\uparrow\mathcal{A}_\downarrow[\Xi_{\mathrm{FFNN}}](\vY^\uparrow, \vY^\downarrow).
\label{eqn:psi_ga}
\end{equation}
This ansatz can be evaluated explicitly with $N_\uparrow!N_\downarrow!$ evaluations of $\Xi_\mathrm{FFNN}$. Importantly, due to the equivariance property of $\vY$, we do not need to antisymmetrize the composed function $\Xi_{\mathrm{FFNN}}\circ \vY$, but only the comparatively small $\Xi_{\mathrm{FFNN}}$. We have found that we can achieve sufficient expressiveness with a very simple choice of $\Xi_{\mathrm{FFNN}}$, further reducing the computational cost. In our experiments, we use a single hidden layer with $N_{\mathrm{FFNN}} = 64$ nodes and a single application of the $\tanh$ activation function, followed by a linear combination operation.  The cost of a single evaluation of $\Xi_\mathrm{FFNN}$ with a single hidden layer with $N_\mathrm{FFNN}$ nodes is $O(d(N_\uparrow + N_\downarrow)N_\mathrm{FFNN})$, resulting in an overall cost of $O(d(N_\uparrow + N_\downarrow)N_\uparrow!N_\downarrow!N_\mathrm{FFNN})$. We depict the construction of the GA layer in \cref{diagram:GA}.

This construction is a universal replacement for the antisymmetry layer in the original FermiNet, due to the universality of neural networks to approximate functions of various desired smoothness classes with an appropriate choice of activation function \cite{Barron1994, Pinkus1999, Rolnick2018, Elbrachter2021}. In fact, the multiplication of two numbers can be approximated to arbitrary accuracy with just four neurons \cite{Lin2017}, and following equations \eqref{eqn:phi_formulation} and \eqref{eqn:determinant_as_antisymmetrized_product}, the FermiNet determinant is simply an antisymmetrized product of simple linear and exponential terms. In this work we choose $\Xi_{\mathrm{FFNN}}$ to be a one-hidden-layer feed-forward neural network for simplicity and efficiency, but one can choose any universal function class, e.g. residual neural networks. 

\subsection{Factorized antisymmetric neural network layer}

\begin{figure}
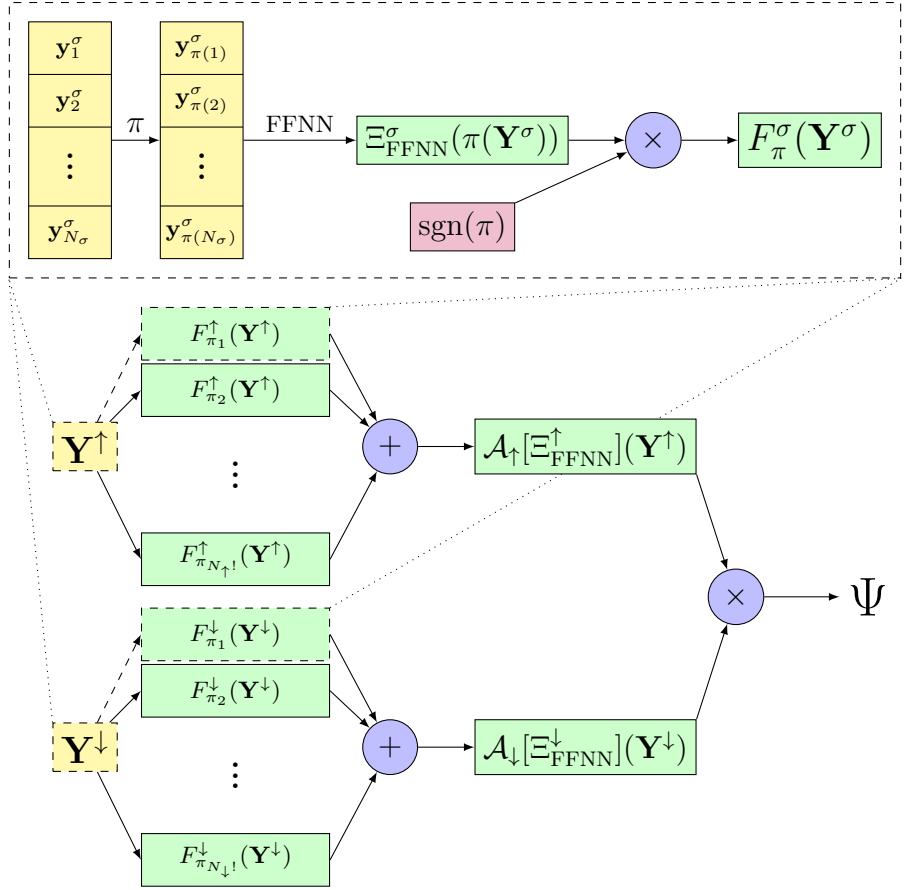

\FAdiagram
\caption{The architecture for the factorized antisymmetric of rank 1 (FA-1). Top: calculation of the wavefunction contribution from a single pseudospin $\sigma$ and permutation $\pi$, denoted by $F^{\sigma}_{\pi}(\vY^\sigma)$. Bottom: combination of contributions for all permutations $\pi$ for both up and down pseudospin components. This is implemented as a batch calculation for each pseudospin, but we show separate arrows for each permutation to emphasize the factorial complexity of the operation. For FA-$K$ this entire computation would be copied $K$ times, with a different feed-forward-neural network in each copy, and the results would be added together.}
\label{diagram:FA}
\end{figure}

Instead of replacing the entire antisymmetric layer of the FermiNet, we can consider replacing each product function $\Xi^\sigma_{\mathrm{FermiNet}}$ in \cref{eqn:determinant_as_antisymmetrized_product} with a feed-forward neural network, arriving at the ansatz
\begin{equation}
\Psi(\vR) = \prod_{\sigma\in\{\uparrow,\downarrow\}} \mathcal{A}_\sigma[\Xi_{\mathrm{FFNN}}^\sigma](\vY^\sigma).
\label{eqn:psi_roa}
\end{equation}
We can relate this ansatz to the generic antisymmetric layer by assuming that $\Xi_{\mathrm{FFNN}}$ admits a functional low rank decomposition
\begin{equation}
\Xi_\mathrm{FFNN}(\vY^{\uparrow}(\vR),\vY^{\downarrow}(\vR)) \approx \sum_{k=1}^K \Xi^{k\uparrow}_\mathrm{FFNN}(\vY^{\uparrow}(\vR)) \Xi^{k\downarrow}_\mathrm{FFNN}(\vY^{\downarrow}(\vR)).
\end{equation}
This is called the factorized antisymmetric layer of rank-$K$ (FA-$K$).
When $K = 1$, we recover the ansatz in \cref{eqn:psi_roa}. We treat this FA-$1$ layer, also depicted in \cref{diagram:FA}, as an especially interesting case given the conjecture of \cite{Hutter2020} that a single generalized Slater determinant may be universal. 
We note that much like in the FermiNet pseudospin determinant terms, the electron positions of different spins do interact with each other in each FA pseudospin term due to the construction of the backflow \eqref{eqn:one_elec_stream_concat}.

We can evaluate the FA-$1$ layer using $N_\sigma!$ evaluations of $\Xi^\sigma_{\mathrm{FFNN}}$ for each pseudospin.  The cost of a single evaluation of $\Xi^\sigma_{\mathrm{FFNN}}$ is $O(dN_\sigma N_{\mathrm{FFNN}})$ operations for the matrix-vector multiplication, and the total cost of the explicit antisymmetrization for FA-$1$ is $O(d(N_\uparrow N_\uparrow!+N_\downarrow N_\downarrow!) N_{\mathrm{FFNN}}).$

It is worth noting that both the factorized and the generic antisymmetric ansatzes have addtional drawbacks beyond the obviously prohibitive factorial scaling. In our experiments with these ansatzes, we observed a great deal of numerical instability due to the massive numerical sign cancellation of the generally non-zero terms in the summations over the symmetric groups. We partially ameliorated this issue by performing the wavefunction evaluation in double precision instead of the more standard single precision (or even half precision) for modern deep learning, but even with this adjustment the numerical stability properties are far too unfavorable to scale these ansatzes as $N_\uparrow$ and $N_\downarrow$ grow large. 
Thus these ansatzes are certainly not intended to be used to directly approximate the ground state wavefunction of heavy atoms or large molecules, but are instead used in this paper as a diagnostic tool for better understanding the empirical performance of the FermiNet backflow and antisymmetry layers. 
More details about the practical effects of the numerical instabilities on our experiments are available in \cref{sec:challenges}.

\subsection{Jastrow factors}

Although the GA architecture can represent general antisymmetric functions on compact domains, we have found that without some mechanism of confining the support size of the wavefunction, the Monte Carlo sampling procedure often becomes unstable. In the standard FermiNet orbitals, this decay is handled by the simple exponential envelope terms in \cref{eqn:phi_formulation}. In our generic antisymmetric layer, however, we have not directly included exponential decay terms in the antisymmetric part, so we require the presence of an additional decay term in the form of a Jastrow factor. We found that including an expressive Jastrow factor greatly increased the stability and accuracy of the ansatz, which suggests that the size of the wavefunction support and the behavior of the tails are of practical importance to the quality of the approximation.

In general, we may implement a Jastrow factor by multiplying an antisymmetric wavefunction ansatz by $\exp\left(J\left(\vR^\uparrow, \vR^\downarrow; \vR^{(a)}\right)\right)$,
where the Jastrow factor $J$ is a function of the electron positions $\vR$ and the nuclei locations $\vR^{(a)}$. $J$ must also be symmetric separately with respect to permutations of $\vR^\uparrow$ and $\vR^\downarrow$ in order to preserve the antisymmetry of the overall wavefunction. In order for $J(\vR)$ to capture the decay of the wavefunction, it will need to satisfy $J\rightarrow-\infty$ as $\abs{\vR}\rightarrow\infty$. The standard Jastrow form~\cite{GubernatisKawashimaWerner2016} which explicitly handles electron-nuclei, electron-electron, and electron-electron-nuclei terms is given by
\begin{equation}
\begin{split}
J(\vR;\vR^{(a)})=& \sum_{\alpha \in\{\uparrow, \downarrow\}} \sum_{i=1}^{N_{\alpha}} \sum_{I=1}^M \chi_{I}\left(\abs{\mathbf{R}_{i}^{\alpha}-\mathbf{R}^{(a)}_{I}}\right)
+ \sum_{\alpha \in\{\uparrow, \downarrow\}} \sum_{\beta \in\{\uparrow, \downarrow\}}\sum_{i=1}^{N_{\alpha}} \sum_{j=1}^{N_{\beta}} u^{\alpha \beta}\left(\abs{\mathbf{R}_{i}^{\alpha}-\mathbf{R}_{j}^{\beta}}\right) \\
+ & 
\sum_{\alpha \in\{\uparrow, \downarrow\}}\sum_{\beta \in\{\uparrow, \downarrow\}}  \sum_{i=1}^{N_{\alpha}} \sum_{j=1}^{N_{\beta}}  \sum_{I} f_{I}^{\alpha \beta}\left(\abs{\mathbf{R}_{i}^{\alpha}-\mathbf{R}_{j}^{\beta}},
 \abs{\mathbf{R}_{i}^{\alpha}-\mathbf{R}^{(a)}_{I}},\abs{\mathbf{R}_{j}^{\beta}-\mathbf{R}^{(a)}_{I}}\right),
\end{split}
\label{eqn:standard_jastrow}
\end{equation}
where the functions $\{\chi_I\}$, $\{u^{\alpha\beta}\}$, and $\{f_I^{\alpha\beta}\}$ satisfy the desired behavior at infinity. One possibility is to use a simple one-body Jastrow from the first term above and let $\chi_I$ represent multiplication by a fixed constant for each nucleus, so that $\chi_{I}(r)=-a_I r$, with $a_I > 0$. Then we have
\begin{equation}
J(\vR;\vR^{(a)})
= \sum_{\alpha \in\{\uparrow, \downarrow\}} \sum_{i=1}^{N_{\alpha}} \sum_{I=1}^M \chi_{I}\left(\abs{\mathbf{R}_{i}^{\alpha}-\mathbf{R}^{(a)}_{I}}\right) 
= \sum_{\alpha \in\{\uparrow, \downarrow\}} \sum_{i=1}^{N_{\alpha}} \sum_{I=1}^M -a_I \abs{\mathbf{R}_{i}^{\alpha}-\mathbf{R}^{(a)}_{I}},
\label{eqn:one_body_jastrow}
\end{equation}
Similarly, one could use the first two terms in \cref{eqn:standard_jastrow} to form a simple two-body Jastrow, with a similar choice for the electron-electron interaction which is identical for the two spin species, i.e. $u^{\alpha\beta}(r) = -\gamma r$, with $\gamma > 0$. These approaches control the support size of the wavefunction, but do not allow much flexibility in the shape of the wavefunction tails outside of the asymptotic regime, where the wavefunction decay is known to be a simple isotropic exponential decay.

To build a more general Jastrow factor, we may leverage the generality of the FermiNet backflow construction to form the backflow-based Jastrow
\begin{equation}
J(\vR;\vR^{(a)}) = -\frac{1}{N}\sum_{\alpha\in\{\uparrow,\downarrow\}}\sum_{i=1}^{N_\alpha}\abs{\vY^\alpha_{\mathrm{Jastrow},i}(\vR)}.
\label{eqn:backflow_jastrow}
\end{equation}
For our numerical results involving the GA and FA-$K$ architectures, we use this general Jastrow expression.
The Jastrow factor needs to be able to grow small as the electron positions move far from the nuclei, which suggests the use of an unbounded activation function.
To achieve this in practice, we simply swap out the tanh activation in \cref{eqn:ferminet_stream_def} for an approximate GeLU activation \cite{Hendrycks2016},
\begin{equation}
\mathrm{GeLU}(x) = 0.5x\left(1 + \tanh\left(\sqrt{\frac{2}{\pi}}\left(x + 0.044715x^3\right)\right)\right)
\end{equation}
which, like the hyperbolic tangent function, has the desirable property of being smooth everywhere.

\subsection{Full determinant FermiNet}

We also explore a variant of the FermiNet called full determinant mode which, like the GA layer, does not assume a factorized form over the two pseudospins. 
In the full determinant mode, the single particle orbitals take exactly the same form as those used in the regular FermiNet architecture. The difference is that instead of using $N_\sigma$ orbitals for each spin $\sigma$, we use $N$ orbitals for both spins, where $N = N_\uparrow + N_\downarrow$. The up- and down-pseudospin orbital matrices are then concatenated into a square $N\times N$ matrix before taking the determinant. The new formula for the orbital matrices is

\begin{equation}
\orbmatfnfull{k}=
\begin{pmatrix}
\orbitalfn{k}{1}{1} & \cdots & \orbitalfn{k}{1}{N_\sigma} \\
\vdots  & \ddots & \vdots \\
\orbitalfn{k}{N}{1} & \cdots & \orbitalfn{k}{N}{N_\sigma} \\
\end{pmatrix},
\end{equation} 
where the only difference from \cref{eqn:orbital_matrix_one_spin} is the change of the maximum orbital index from the pseudospin-specific $N_\sigma$ to the total particle count $N$. Therefore $\orbmatfnfull{k}$ is a matrix of size $N\times N_{\sigma}$.
The final ansatz is then generated as the sum of the determinants of the concatenated orbital matrices: 
\begin{equation} 
\Psi_{\mathrm{FermiNet},\mathrm{full}}(\vR) = \sum_{k=1}^K \det \left[  \orbmatfull[\uparrow]{k} , \, \orbmatfull[\downarrow]{k} \right].
\end{equation}

The idea behind this construction is to provide a more flexible way to treat the interactions between the two pseudospin components. Importantly, because $\vY$ is only equivariant with respect to permutations which exchange particles of the same spin, the concatenated determinant does not enforce an antisymmetry constraint between particles of opposite spins. We also note that it is possible to reconstruct the original FermiNet ansatz as a special case of the full determinant ansatz, by setting $\orbitalfn{k}{i}{j} = 0$ whenever $\sigma = {\uparrow}$ and $i > N_\uparrow$ or $\sigma = {\downarrow}$ and $i \le N_\uparrow$. In that case the full matrix becomes block-diagonal and the determinant factors into a simple product of pseudospin determinants~\cite{Pfau2020}. 
We are particularly interested in the evaluating the performance of this full determinant mode when $K=1$, which we refer to as the full single-determinant FermiNet.

\section{Optimization}
\newcommand{\loss}{\mathcal{L}(\theta)}
\newcommand{\intR}[1]{\int #1 \,d\vR}
\newcommand{\psiR}{\Psi_\theta(\vR)}
\newcommand{\psiRC}{\Psi_\theta^*(\vR)}
\newcommand{\psiX}{\Psi_\theta(\vX)}
\newcommand{\psiXC}{\Psi_\theta^*(\vX)}
\newcommand{\emploss}{\widetilde{\mathcal{L}}(\theta)}
\newcommand{\expect}[1]{\EE_{p(\vR)} \left[ #1 \right]}
\newcommand{\xxt}[1]{#1#1^\intercal}
\newcommand{\del}[1]{\partial #1}
\newcommand{\dlogpsi}{\del{\log \abs{\psiR}}}

\subsection{Gradient calculation}
When computing parameter updates, estimating the gradient of $\mathcal{L}(\theta)$ by directly differentiating the empirical risk $\tilde{\mathcal{L}}(\theta)$ defined in \cref{eqn:loss_def} using an automatic differentiation framework is generally difficult due to the dependence of the Monte Carlo sampling on the parameters $\theta$. However, the following standard unbiased estimate of the gradient of the true expected energy $E[\Psi_\theta] \equiv \mathcal{L}(\theta)$ is available for real wavefunctions \cite{Bressanini1999}:
\begin{equation}
\begin{split}
\partial_\theta \loss &=\frac{\intR{2(\partial_\theta\log\abs{\Psi_{\theta}})(E_L(\vR;\theta) - \mathcal{L}(\theta))\abs{\Psi_{\theta}}^2}}{\intR{ \abs{\Psi_{\theta}}^2}} \\
&\approx \frac{1}{|\xi_{\theta}|}\sum_{\vR\in\xi_\theta} 2(\partial_\theta\log\abs{\Psi_\theta})(E_L(\vR;\theta) - \tilde{\mathcal{L}}(\theta))
\end{split}
\end{equation}
where $\xi_\theta$ are a set of samples from the density $p_\theta(\vR) = |\Psi_\theta(\vR)|^2 / \intR{ \abs{\Psi_{\theta}}^2}$. For completeness, the derivation of the gradient is provided in Appendix \ref{section:gradient}.

The zero-variance principle \cite{Coldwell1977} states that the eigenstates of the Hamiltonian (\cref{eqn:manybody_ham}) will have the same local energy everywhere. This improves the quality of the loss and gradient approximations as the training converges. In addition, this principle can make the variance of the local energy an attractive target for minimization. Indeed, variance minimization has an extensive history in the quantum Monte Carlo space \cite{Umrigar1988, Kent1999, Umrigar2005}. Nonetheless, we follow the work of FermiNet \cite{Pfau2020, Spencer2020} and only use energy minimization to optimize our wavefunctions.

\subsection{Optimizer}
The choice of efficient optimization algorithms for parameter updates in variational Monte Carlo has historically been a complex issue and is still under active debate (see e.g.~\cite{UmrigarFilippi2005,NeuscammanUmrigarChan2012,OtisNeuscamman2019,Becca2017,Carleo2017,SabzevariMahajanSharma2020,StokesRobledo2020,Pfau2020}). Among these works, Ref.~\cite{Pfau2020} provided  evidence that the use of the Kronecker Factorized Approximate Curvature (KFAC) method \cite{Martens2015} can be advantageous when compared to standard stochastic gradient descent-like methods used in the machine learning community such as Adam \cite{kingma2017}. KFAC is a method for approximating natural gradient descent efficiently by preconditioning the gradient with an approximate inverse of the Fisher information matrix. Both the overall structure of KFAC and the extra steps required to apply KFAC to an unnormalized wavefunction are described succinctly in \cite{Pfau2020}. For the convenience of the reader we reproduce here an overview of these topics.

In the exact natural gradient descent method, the gradient of the loss function is multiplied by the inverse of the Fisher information matrix before using the gradient to make a parameter update \cite{Amari1998}. This has the effect of taking the path of steepest descent not in Euclidean parameter space, but in the space of probability distributions defined by the model, with distance measured by the KL-divergence \cite{Amari2000}. Concretely, updates in natural gradient descent take the form
\begin{equation}
\theta' = \theta - \eta \mathcal{F}^{-1} \nabla_\theta \loss.
\end{equation}
Here $\eta \in \RR$ is the learning rate and $\mathcal{F}$ is the Fisher information matrix defined as
\begin{equation}
\mathcal{F}_{ij} = \expect{ \frac{\partial \log p(\vR)}{\partial \theta_i} \frac{\partial \log p(\vR)}{\partial \theta_j}}.
\end{equation}
Note that in our case $p(\vec{R}) \propto  |\psiR|^2 $, though the two are not equal as $\Psi$ is not necessarily normalized. In fact, obtaining the Fisher information matrix requires a slightly different calculation in an unnormalized setting, which is usually referred to as stochastic reconfiguration \cite{Becca2017}:
\begin{equation}
\mathcal{F}_{ij} \propto \expect{\left( \mathcal{O}_i - \expect {\mathcal{O}_i } \right) \left( \mathcal{O}_j - \expect{ \mathcal{O}_j }\right)}, 
\end{equation}
where $ \mathcal{O}_i = \partial \log \abs{\psiR}/\partial \theta_i$. The equivalence of this formulation is proved in Appendix C of \cite{Pfau2020}. In the setting of quantum information geometry, the Fisher information matrix is proportional to, and perhaps more accurately viewed as, the Fubini-Study metric tensor or quantum geometric tensor \cite{StokesIzaac2020}.

Directly inverting the Fisher information matrix is infeasible for large models, as the matrix dimensions scale directly with number of parameters. KFAC solves this problem by making two approximations to the Fisher matrix to allow its efficient inversion. The first is to assume that the Fisher entries for weights in different layers of the network are zero. This assumption reduces the Fisher matrix to a block diagonal form, so that inverting the remaining matrix only requires inverting each block independently. The second is based on the observation that the block corresponding to each layer of the network can be written as the mean-centered covariance of a Kronecker product of two vectors, one consisting of neuron activation values for the inputs to the layer and the other consisting of gradients of the loss with respect to the outputs of the layer. KFAC replaces this with the Kronecker product of the mean-centered covariance of the same vectors. As discussed in \cite{Martens2015}, this is a significant and theoretically unsupported approximation, but seems to work well in practice, at least in some use cases.

In our experiments, we rely on the JAX implementation of KFAC provided by the work of \cite{Pfau2020}. In using this implementation, we register all dense layers in our networks with KFAC, including those within the feed-forward neural networks of our generic antisymmetric and factorized antisymmetric layers. This ensures that we use the Kronecker product approximation of the Fisher matrix for all layers in the network, rather than defaulting to a simpler diagonal approximation.

\section{Numerical Experiments}
\label{section:results}
In this section, we compare the previously described architectures on small atomic and molecular systems. All numerical experiments with the factorized and generic antisymmetric layers are performed using the VMCNet repository~\cite{vmcnet2021github}, which is based on the JAX framework \cite{jax2018github}. In \cref{section:code_benchmarking}, we demonstrate the comparability of the VMCNet repository with the JAX branch of the FermiNet repository~\cite{Spencer2020}. Using JAX allows us to leverage the implementation of KFAC in~\footnote{\url{https://github.com/deepmind/deepmind-research/tree/master/kfac_ferminet_alpha}}, take advantage of the flexibility provided by JAX's clean functional style, and enjoy the performance benefits granted by its excellent out-of-the-box GPU utilization and just-in-time compilation. We used A100 GPUs on the Google Cloud Platform (GCP) for any calculations that required double precision, and GTX 2080TI GPUs with the Berkeley Research Computing (BRC) program for all other calculations.

To estimate energy values accurately after training, we ran pure MCMC for a large number of iterations without performing parameter updates, collecting samples every 10 iterations. We also estimated the integrated autocorrelation of the local energy during these evaluation runs in order to get a robust estimate of the standard error of our energy estimates. The hyperparameters we used, including the number of training and evaluation iterations, are listed in Appendix \ref{section:hyperparams}. The gradient clipping and sampling procedures are described in Appendix \ref{section:sampling_and_clipping}.

Throughout this section, we use the following standard notation to present our numerical results. All units are atomic units (a.u.) unless otherwise specified. The estimator of the energy used is the sample mean followed by, in parentheses, the standard error in the last digit(s) of the estimate. For example, -54.58868(4) means a sample mean of -54.58868 a.u. with a standard error of approximately $4\times 10^{-5}$ a.u., and -75.06314(13) means a sample mean of -75.06314 a.u. with a standard error of approximately $1.3\times 10^{-4}$ a.u.. The error of the energy is also measured by the percentage of the correlation energy recovered. The correlation energy is defined to be the difference between the Hartree-Fock energy and the exact ground state energy, so that recovering $0\%$ of the correlation energy means that the calculation produces the Hartree-Fock energy, and recovering $100\%$ means the calculation is exact. The correlation energy itself only contributes a tiny amount, usually less than $1\%$, to the ground state total energy, but capturing the correlation energy accurately is extremely important in chemistry.

\subsection{Performance: atomic systems}

\begin{figure}
    \centering
    \includegraphics[width=\textwidth]{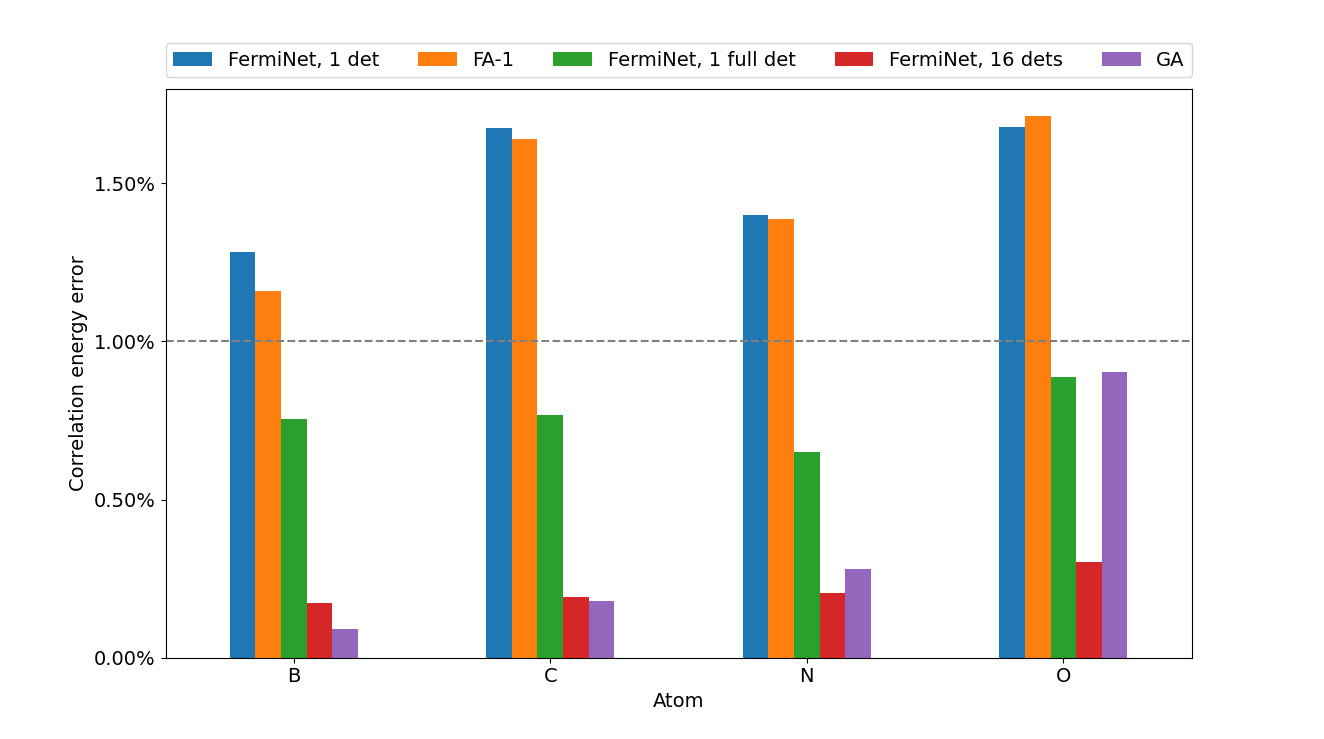}
    \caption{Comparison of methods on atomic systems. FA stands for factorized antisymmetric layer, and GA stands for generic antisymmetric layer, as discussed in section \ref{section:architecture}. The 16 determinant FermiNet numbers are taken from Ref.~\cite{Pfau2020}. The GA result on the oxygen atom uses the more restrictive one-body Jastrow, and the parameters may not be fully optimized due to the limitations of our resources. Dashed line indicates $1\%$ of the error in correlation energy.}
    \label{fig:atom_compare}
\end{figure}

\begin{table}
\centering
\caption{Comparison of methods on atomic systems. FA stands for factorized antisymmetric layer, and GA stands for generic antisymmetric layer, as discussed in section \ref{section:architecture}. The 16 determinant FermiNet and Hartree-Fock results are taken from Ref.~\cite{Pfau2020}.}
\label{tab:atom_compare}
\begin{tabular}{@{}lccccccc@{}}
\toprule
\multicolumn{1}{c}{} & \begin{tabular}[c]{@{}c@{}}FermiNet\\ 1 det\end{tabular} & FA-1                                                & \begin{tabular}[c]{@{}c@{}}FermiNet\\ 1 full det\end{tabular} & \begin{tabular}[c]{@{}c@{}}FermiNet\\ 16 dets  \cite{Pfau2020}\end{tabular} & GA                                                                                                                                                     & HF \cite{Pfau2020}         & Reference \cite{Chakravorty1993} \\ \midrule
B                    & -24.65236(3)                                             & -24.65251(2)                                        & -24.65300(3)                                                  & -24.65370(3)                                                             & -24.65380(2)                                                                                                                                           & \multirow{2}{*}{-24.53316} & \multirow{2}{*}{-24.65391}       \\
corr \%              & 98.71(2)\%                                               & 98.84(2)\%                                          & 99.25(2)\%                                                    & 99.83(3)\%                                                               & 99.91(1)\%                                                                                                                                             &                            &                                  \\ \midrule
C                    & -37.84247(4)                                             & -37.84252(3)                                        & -37.84384(4)                                                  & -37.84471(5)                                                             & -37.84473(3)                                                                                                                                           & \multirow{2}{*}{-37.6938}  & \multirow{2}{*}{-37.8450}        \\
corr \%              & 98.33(3)\%                                               & 98.36(2)\%                                          & 99.23(3)\%                                                    & 99.81(3)\%                                                               & 99.82(2)\%                                                                                                                                             &                            &                                  \\ \midrule
N                    & -54.58662(5)                                             & -54.58664(4)                                        & -54.58800(8)                                                  & -54.58882(6)                                                             & -54.58868(4)                                                                                                                                           & \multirow{2}{*}{-54.4047}  & \multirow{2}{*}{-54.5892}        \\
corr \%              & 98.60(3)\%                                               & 98.61(2)\%                                          & 99.35(5)\%                                                    & 99.79(3)\%                                                               & 99.72(4)\%                                                                                                                                             &                            &                                  \\ \midrule
O                    & -75.06314(13)                                            & -75.06305(6)                                        & -75.06510(6)                                                  & -75.06655(7)                                                             & -75.06506(6)\footnote{Due to the limitations of our computational resources, this result uses the simple but more restrictive one-body Jastrow from \cref{eqn:one_body_jastrow}, and the parameters may not be fully optimized.} & \multirow{2}{*}{-74.8192}  & \multirow{2}{*}{-75.0673}        \\
corr \%              & 98.32(5)\%                                               & 98.29(3)\%                                          & 99.11(2)\%                                                    & 99.70(3)\%                                                               & 99.10(3)\%                                                                                                                                             &                            &                                  \\ \bottomrule
\end{tabular}
\end{table}

We test our generic and factorized antisymmetric architectures on a few small atoms with nuclear charge from five to eight and compare these results to the results of FermiNet with 1 determinant, FermiNet with 1 full determinant, and FermiNet with 16 determinants. In Table \ref{tab:atom_compare}, we compare the attained energies on these architectures after training with KFAC. These results are depicted as well in Figure \ref{fig:atom_compare}.

We find that the generic antisymmetric layer attains highly accurate energies when paired with the backflow-based Jastrow, achieving greater than 99.7\% of the correlation energy. For the smallest systems, i.e. boron and carbon, the FermiNet-GA ansatz does at least as well as many-determinant FermiNet. For the larger systems, the performance of FermiNet-GA in our implementation began to suffer noticeably as we hit the limitations of our computational resources. For example, our result on oxygen for the generic antisymmetric architecture used only the simple one-body Jastrow in \cref{eqn:one_body_jastrow} and may not have reached the lowest energy that could be attained with additional training. We provide further discussion of the challenges with numerical stability and computational cost in \cref{sec:challenges}.

\begin{figure}
\centering
\includegraphics[width=0.8\textwidth]{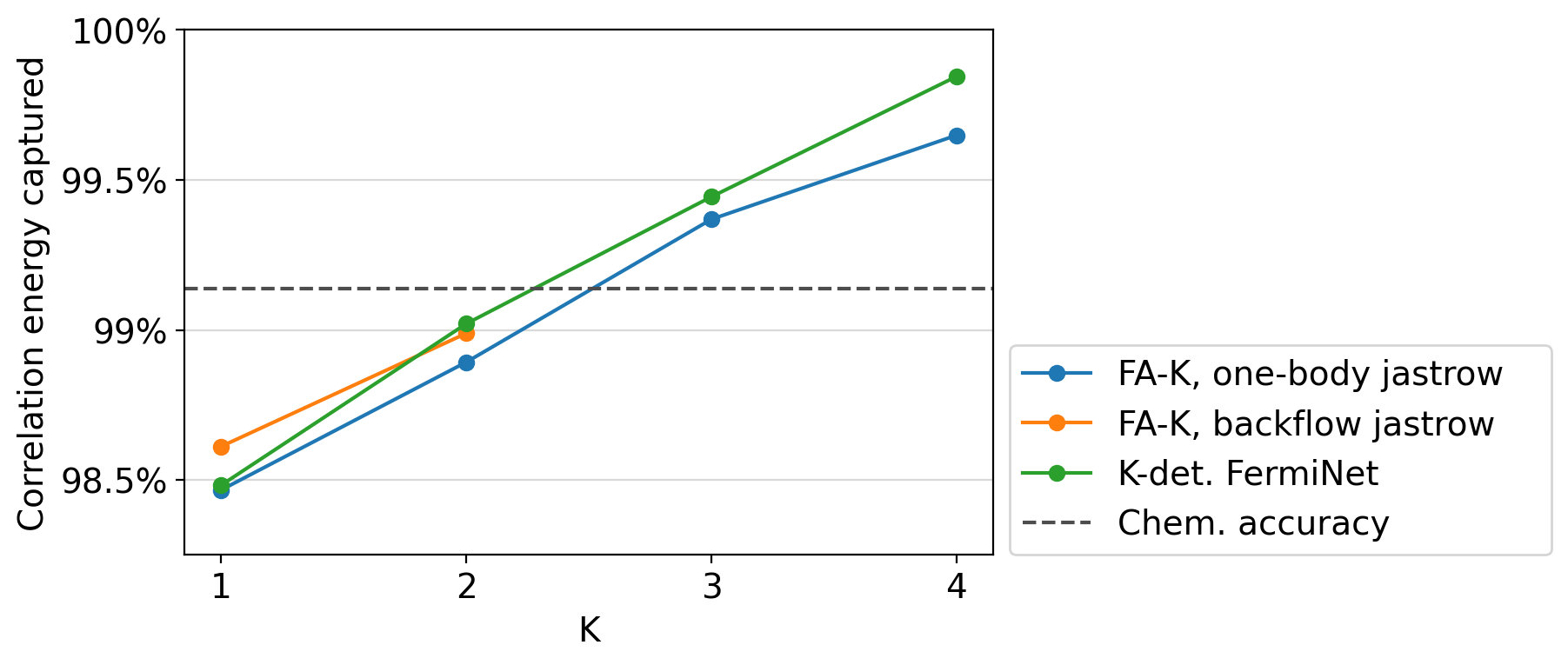}
\caption{Comparison of factorized antisymmetric layer of rank $K$ with $K$-determinant FermiNet for $K=1,2,3,4$ on the nitrogen atom. Data for FA-$K$ is presented with both the simple one-body Jastrow and the more expressive backflow based Jastrow, though results for the backflow Jastrow are limited to $K\le2$ due to numerical stability issues and computational resource constraints. Data for multi-determinant FermiNet represent the best of several runs to account for run-to-run variance.}
\label{fig:rank_k}
\end{figure}

Interestingly, we do not see a gap in the attained energy between the factorized antisymmetric layer of rank 1 and single-determinant FermiNet for any system except for boron. To explore this trend further, we compare in Figure \ref{fig:rank_k} the factorized antisymmetric layers of rank 1 through 4 against the FermiNet with 1 through 4 determinants, all on the nitrogen atom. We find that FA-$K$ performs approximately equivalently to $k$-determinant FermiNet in all cases, and in most cases it performs slightly worse. This comparison is telling since replacing each generalized Slater determinant in the $K$-determinant FermiNet with an explicitly antisymmetrized feedforward neural network yields exactly the FA-$K$ architecture. The fact that this does not yield a performance improvement suggests that the reason $K$-determinant FermiNet is not fully general is not due to the structure of the individual generalized Slater determinants, but rather due to the sum of products structure that is used to combine the generalized Slater determinants together. This appears to be true even though the sum of products is only taken with respect to the pseudospin components generated by the backflow rather than the original spins.

\begin{figure}
\centering
    \begin{subfigure}[b]{0.45\textwidth}
        \includegraphics[width=\textwidth]{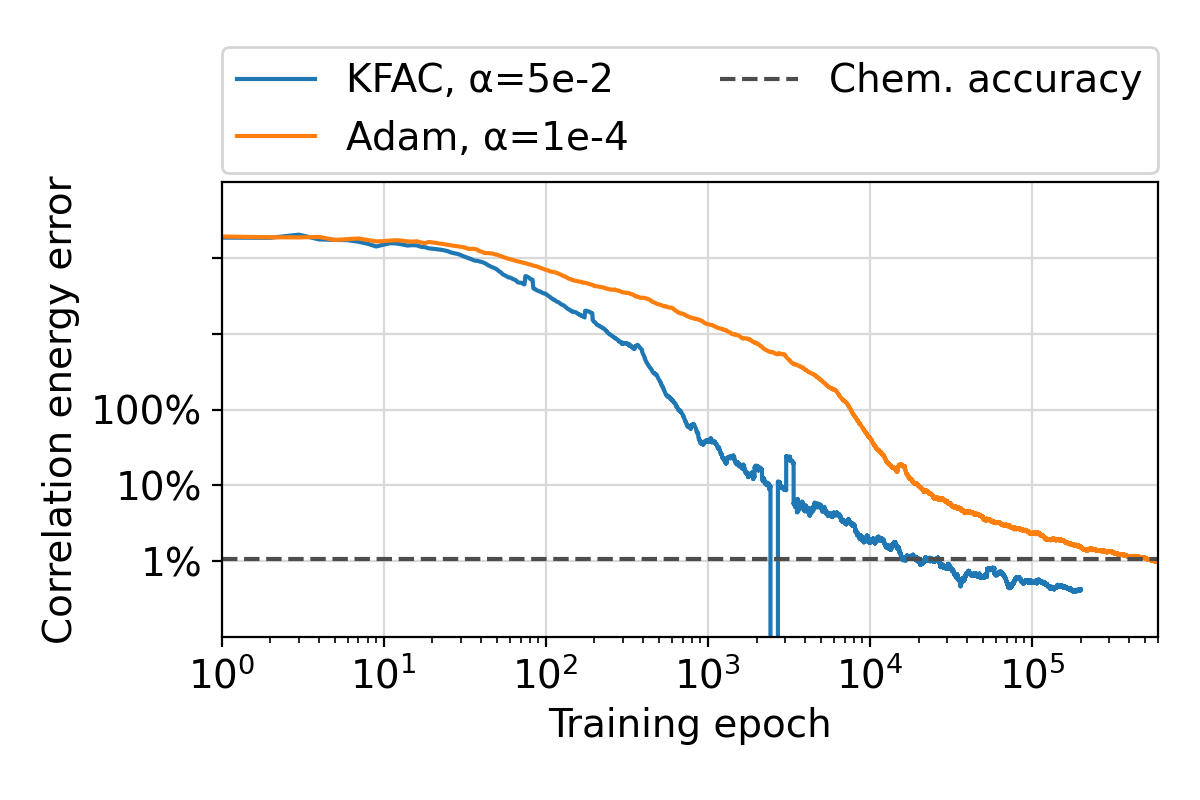}
        \caption{Generic antisymmetry}
    \end{subfigure}
    \begin{subfigure}[b]{0.45\textwidth}
        \includegraphics[width=\textwidth]{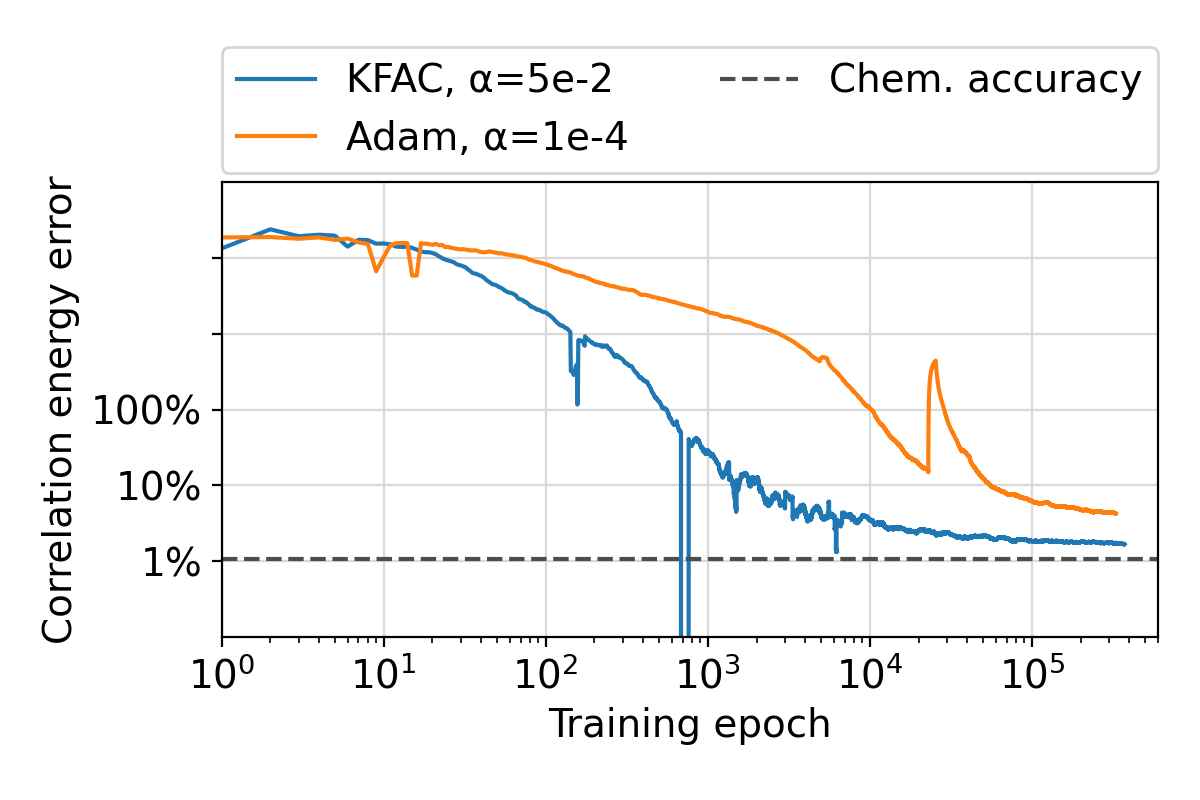}
        \caption{Factorized antisymmetric layer of rank 1}
    \end{subfigure}
\caption{Adam versus KFAC for the optimization of the generic and factorized antisymmetry architectures on the carbon atom. The proposed updates prior to the application of the learning rate may have different scales for the two optimizers, so we choose the largest stable initial learning rate for each from a coarse sweep of learning rates $\alpha$ with $\log_{10}\alpha \in [-4, -1]$. At each epoch, rolling averages of the previous 10\% of training epochs are shown here for clarity. One epoch is one parameter update. }
\label{fig:kfac_vs_adam}
\end{figure}

\begin{figure}
\centering
    \begin{subfigure}[b]{0.45\textwidth}
        \includegraphics[width=\textwidth]{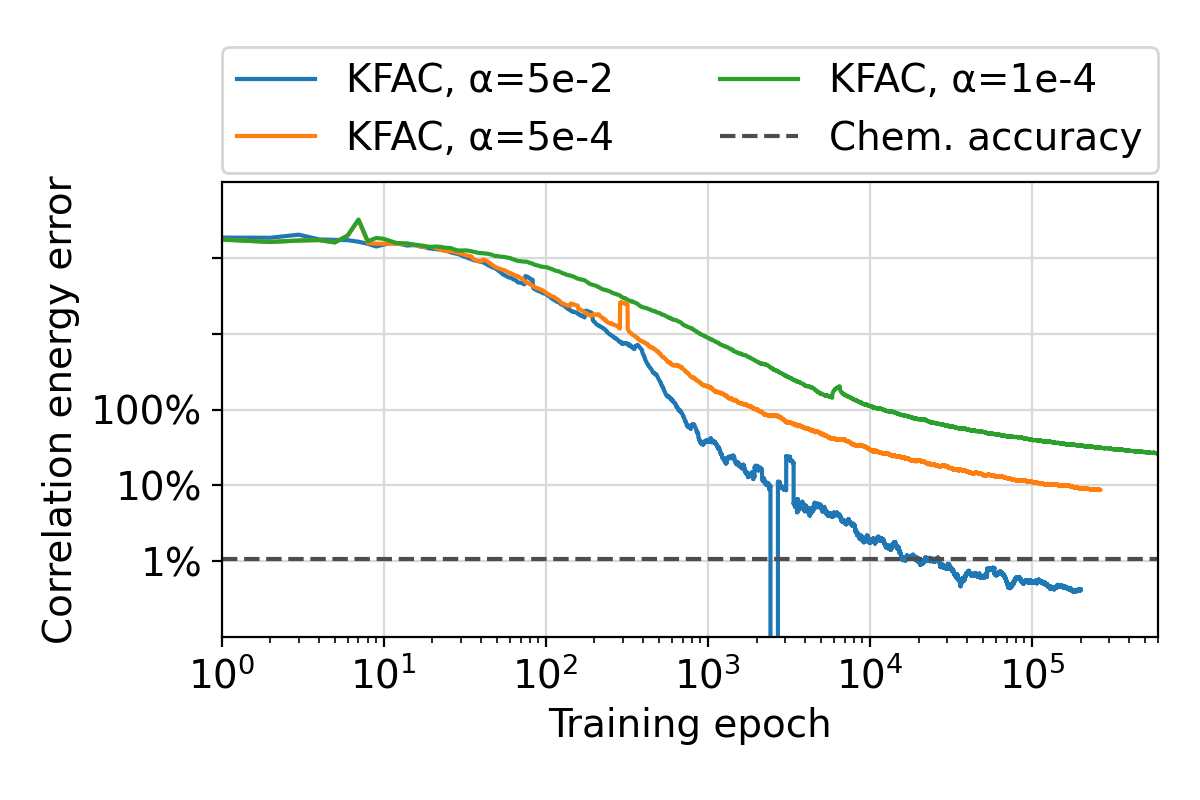}
        \caption{Generic antisymmetry}
    \end{subfigure}
    \begin{subfigure}[b]{0.45\textwidth}
        \includegraphics[width=\textwidth]{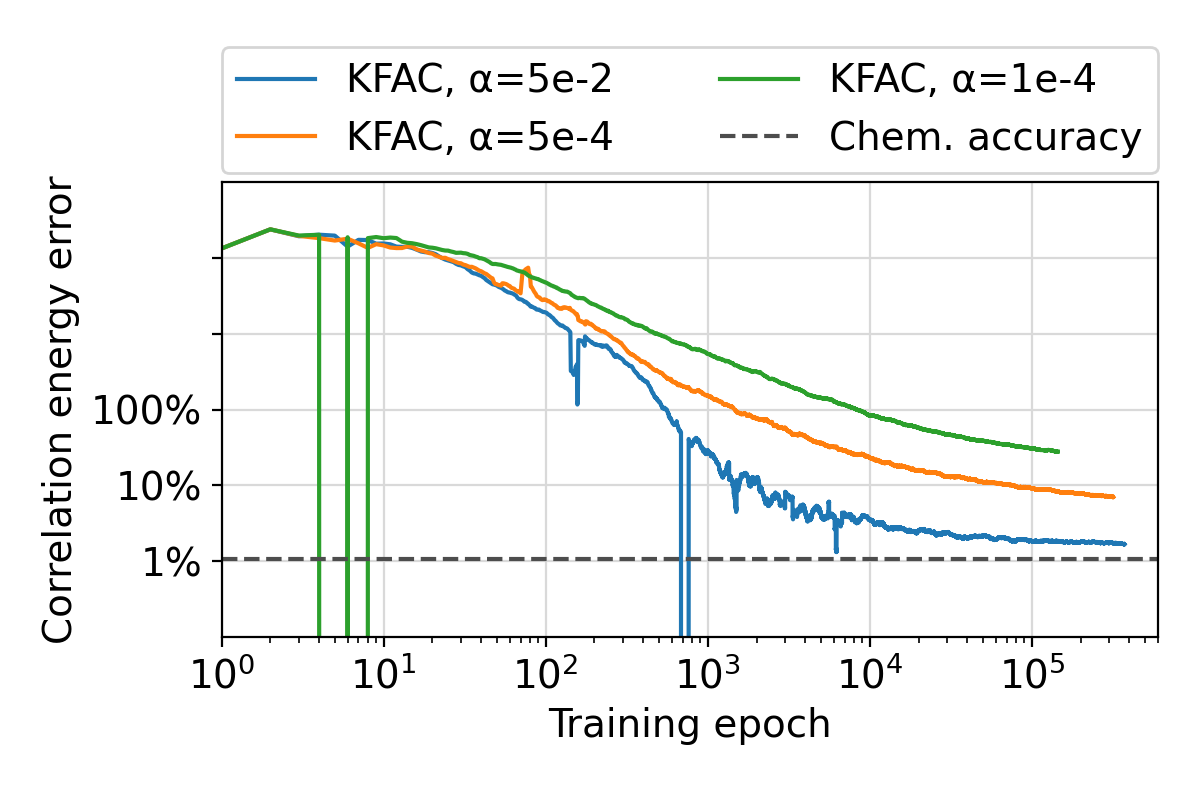}
        \caption{Factorized antisymmetric layer of rank 1}
    \end{subfigure}
\caption{KFAC at a few learning rates $\alpha$ for the optimization of the generic and factorized antisymmetry architectures on the carbon atom. At each epoch, rolling averages of the previous 10\% of training epochs are shown here for clarity. One epoch is one parameter update. There is occasionally some initial instability within the first 10 or so epochs.}
\label{fig:kfac_learning_rates}
\end{figure}

We also provide additional evidence in favor of the claim in \cite{Pfau2020} that KFAC provides an advantage over Adam when optimizing FermiNet-like architectures for small atoms and molecules. In Figure \ref{fig:kfac_vs_adam} we provide a log-log plot of the correlation energy error during training of the generic and factorized antisymmetric architectures on the carbon atom. In this figure, the learning rate schedule for KFAC was chosen to be $5\cdot 10^{-2} / (1 + 10^{-4}t)$. The learning rate schedule differed slightly for Adam, chosen instead to be $10^{-4} / (1 + 10^{-4}t)$. To determine the initial learning rate for both Adam and KFAC, we coarsely swept over a range of initial learning rates between 1e-4 and 1e-1 on the Carbon atom and picked the learning rate which resulted in the lowest final energies without encountering numerical instability or NaNs. The difference in the learning rates which we found were best for the two optimizers may be due to the different scale of the updates prior to the learning rate scaling. Figure \ref{fig:kfac_learning_rates} shows that the choice of learning rate is quite important, and we generally found that in our experiments, using the highest consistently stable initial learning rate resulted in the lowest energies overall. When the learning rate is chosen in this way, KFAC reaches similar energy levels to Adam with as many as two orders of magnitude fewer epochs.

\subsection{Performance: square H$_4$ model}

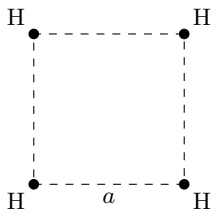
\begin{figure}
\begin{tikzpicture}

\coordinate (1) at (0,0);
\coordinate (2) at (2,0);
\coordinate (3) at (2,2);
\coordinate (4) at (0,2);
\coordinate (5) at ($(1)!.5!(2)$);

\fill (1) circle (2pt) node [below left] {H};
\fill (2) circle (2pt) node [below right] {H};
\fill (3) circle (2pt) node [above right] {H};
\fill (4) circle (2pt) node [above left] {H};
\node at (5) [below] {$a$};

\draw[dashed] (1)--(2)--(3)--(4)-- cycle;

\end{tikzpicture}
\caption{Atomic configuration for the square H$_4$ model. The side length is $a$.}
\label{fig:H4_setup}
\end{figure}

\begin{figure}
    \centering
    \begin{subfigure}[b]{0.45\textwidth}
        \includegraphics[width=\textwidth]{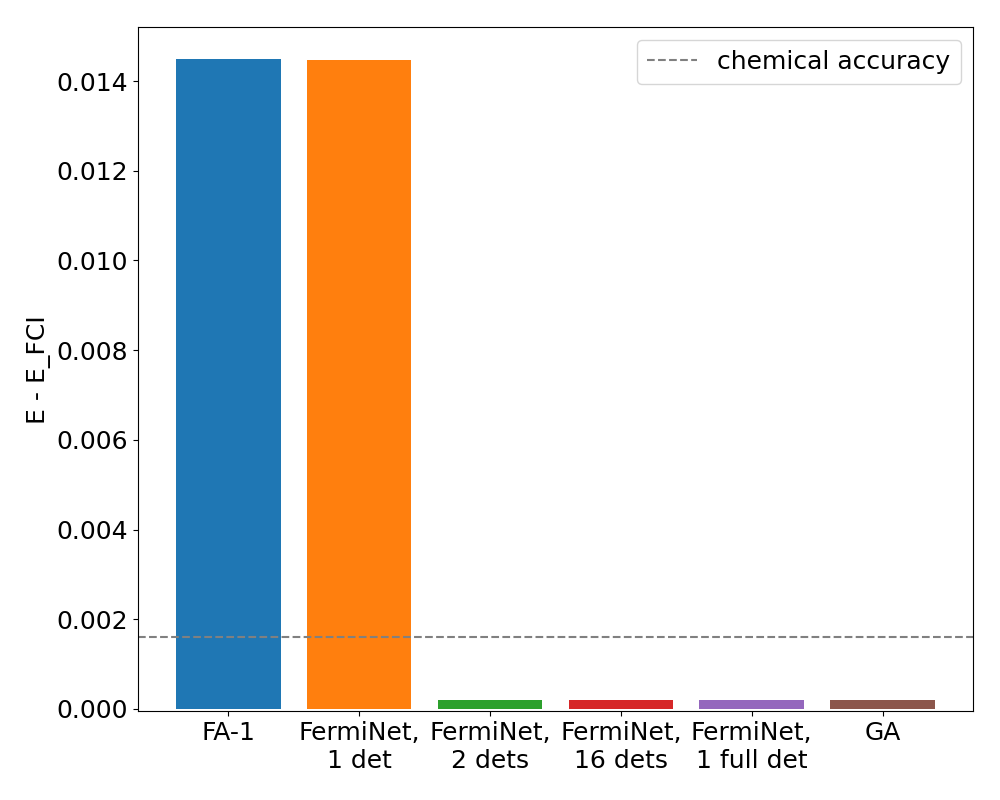}
        \caption{$a = 1.0$ Bohr}
    \end{subfigure}
    \begin{subfigure}[b]{0.45\textwidth}
        \includegraphics[width=\textwidth]{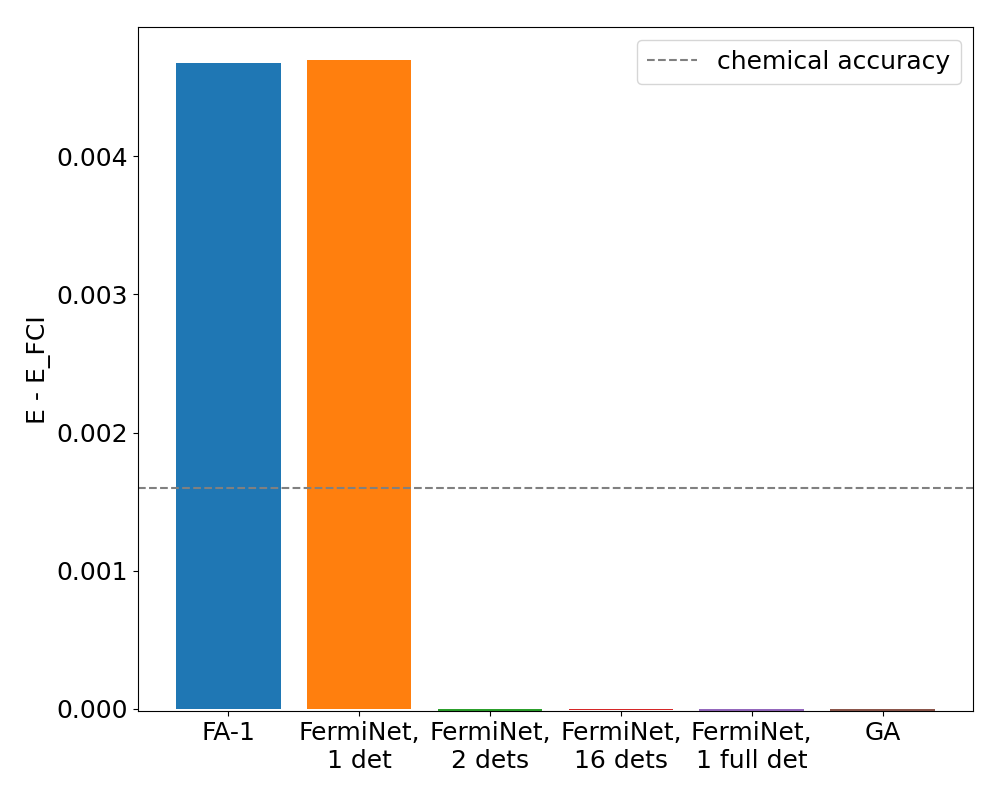}
        \caption{$a = 4.0$ Bohr}
    \end{subfigure}
    \caption{Error of the total energy attained by various VMC ansatzes on the square H$_4$ model of different bond lengths. Details on the complete basis set extrapolation for FCI are included in \cref{sec:basis_set_extrap}.}
    \label{fig:H4_compare}
\end{figure}

\newcolumntype{C}[1]{>{\centering\let\newline\\\arraybackslash\hspace{0pt}}m{#1}}
\begin{table}
\centering
\caption{Comparison of methods on the square H$_4$ model. Details on the complete basis set extrapolation for RHF/UHF/FCI are included in \cref{sec:basis_set_extrap}.}
\label{tab:H4_compare}
\begin{tabular}{@{}C{4em}C{7em}C{7em}C{7em}C{7em}C{7em}@{}}
\toprule
\multicolumn{1}{l}{$a$ (Bohr)} & RHF       & UHF       & FA-1                                                & \begin{tabular}[c]{@{}c@{}}FermiNet\\ 1 det\end{tabular} & FCI     \\ \midrule
1.0                            & -1.2863   & -1.3358   & -1.424518(7)                                        & -1.424531(6)                                             & -1.4390 \\
4.0                            & -1.7492   & -2.0092   & -2.029525(4)                                        & -2.029502(4)                                             & -2.0342 \\ \toprule
\multicolumn{1}{l}{$a$ (Bohr)} & \begin{tabular}[c]{@{}c@{}}FermiNet\\ 2 dets\end{tabular} & \begin{tabular}[c]{@{}c@{}}FermiNet\\ 16 dets\end{tabular} & \begin{tabular}[c]{@{}c@{}}FermiNet\\ 1 full det\end{tabular} & GA           & FCI     \\ \midrule
1.0                            & -1.438804(5)                                              & -1.438796(5)                                               & -1.438805(4)                                                  & -1.438799(4) & -1.4390 \\
4.0                            & -2.034212(3)                                              &       -2.034208(3)                                         & -2.034217(2)                                                  & -2.034218(3) & -2.0342 \\
\bottomrule
\end{tabular}
\end{table}

The square H$_4$ model (\cref{fig:H4_setup}) provides an interesting case study as a prototypical strongly correlated system~\cite{JankowskiPaldus1980}.
Unlike the atomic case, the simple product of a pseudospin-up and pseudospin-down antisymmetry is inadequate to capture the ground state within a few percent of the correlation energy. In \cref{fig:H4_compare}, we see that in both FA-1 and the standard single-determinant FermiNet, the energy attained was significantly higher than that of any of the other neural network ansatzes tested here.  We find excellent agreement between multiple-determinant FermiNet, full single-determinant FermiNet, and FermiNet-GA, agreeing within the estimated Monte-Carlo error. The failure of FA-1 to capture more of the ground-state energy than the standard single-determinant FermiNet again suggests that, at least for the small atomic and molecular systems modeled here, the FermiNet architecture is already very expressive for each pseudospin antisymmetry, even without an explicit Jastrow factor. The fact that even the addition of the general backflow-based Jastrow to the FA-1 architecture does not yield better results than FermiNet suggests that the lack of expressiveness of these simple ``rank-one'' product wavefunction ansatzes has to do with their nodal structure.

\subsection{Comparison of nodal surfaces}
\newcommand{\nodalpictureli}[3]{\includegraphics[valign=m,width=#1]{img/nodes/lithium/#2_-5.0_5.0_150_#3_plot.png}}
\begin{figure}
\centering
\includegraphics[width=0.8\textwidth]{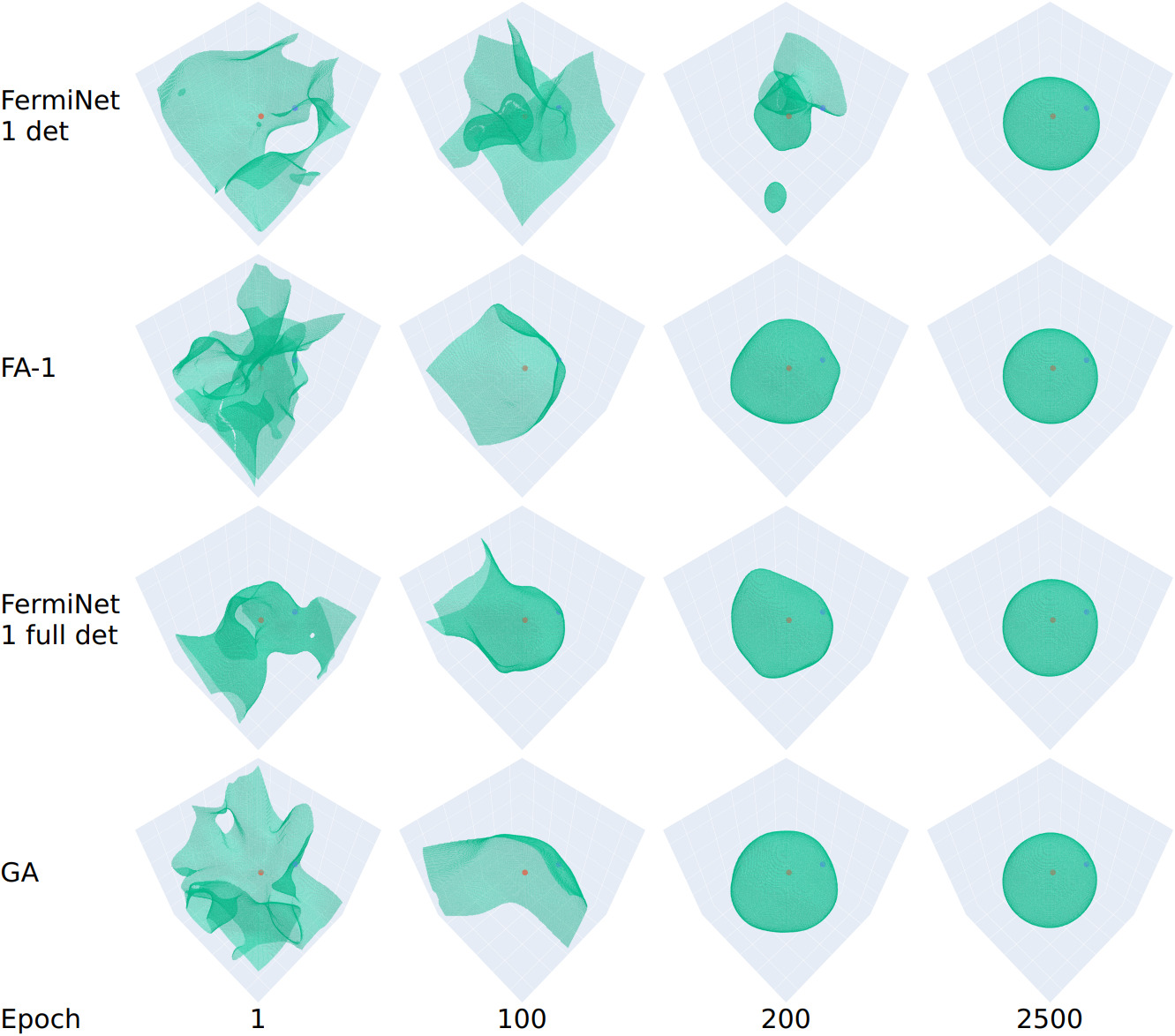}
\caption{Nodal surface cross-sections of various ansatzes on the lithium atom during training. The locations of two electrons are fixed, with one spin-up (same spin) electron in blue and one spin-down (opposite spin) electron in red. The fixed electrons are in random locations. The surfaces shown are produced by evaluating the sign of the wavefunction on the position of the remaining (spin-up) electron in the box $[-5,5]^3$ on 150 points in each direction and using the Isosurface graph object of the Plotly python graphing library.}
\label{fig:nodal_surfaces_li}
\end{figure}

\newcommand{\nodalpicturebe}[3]{\includegraphics[valign=m,width=#1]{img/nodes/beryllium/#2_-5.0_5.0_150_#3_plot.png}}
\begin{figure}
\centering
\includegraphics[width=0.8\textwidth]{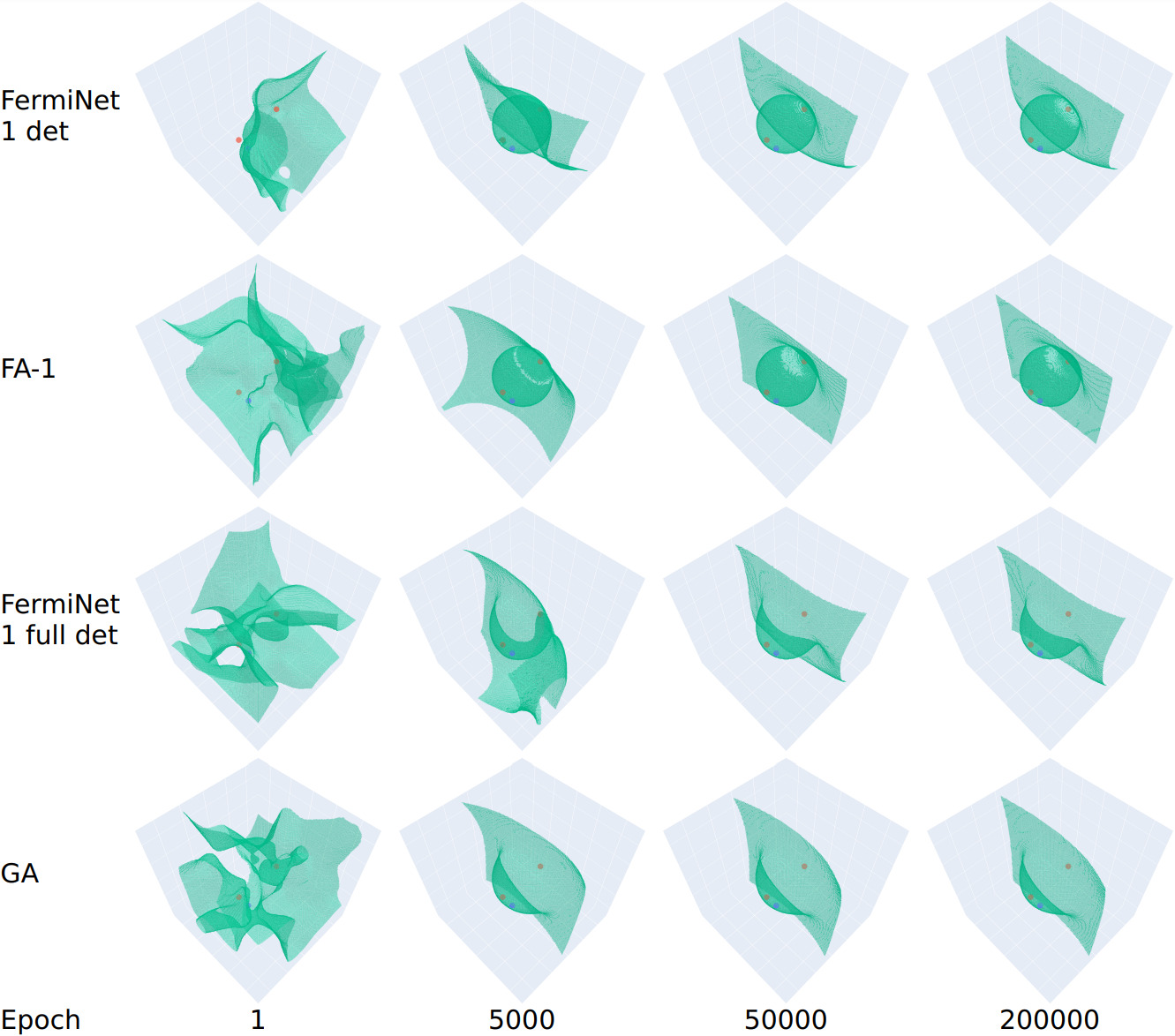}
\caption{Nodal surface cross-sections of various ansatzes on the beryliium atom during training. The locations of three electrons are fixed, with one spin-up (same spin) electron in blue and two spin-down (opposite spin) electron in red. The fixed spin-up electron is placed at $(2, 1, 0)$, and the two spin-down electrons are placed at $(0, \pm 2, 0)$. The surfaces shown are produced by evaluating the sign of the wavefunction on the position of the remaining (spin-up) electron in the box $[-5,5]^3$ on 150 points in each direction and using the Isosurface graph object of the Plotly python graphing library.}
\label{fig:nodal_surfaces_be}
\end{figure}

Given a sufficiently general Jastrow correlation factor, the essential difficulty in the expressiveness of trial wavefunctions for quantum Monte Carlo methods lies in the accurate modeling of the nodal hypersurface~\cite{Ceperley1991}. We thus explore the nodal hypersurfaces generated by several of our ansatzes in Figures \ref{fig:nodal_surfaces_li} and \ref{fig:nodal_surfaces_be}, taking inspiration from \cite{Ceperley1991}. In these figures, we fix the locations of all but one electron in the lithium and beryllium atoms and plot the nodal surface  of the resulting one-body functions in the final electron position for four of our ansatzes: standard single-determinant FermiNet, FA-1, full single-determinant FermiNet, and GA. This plotted nodal surface is thus a 3-dimensional cross-section of the full $(3N - 1)$-dimensional nodal hypersurface of the many-body wavefunction, where for example for the beryllium atom $3N-1 = 11$.

The nodal surface of the lithium wavefunction is essentially described by the two-particle antisymmetry between the two electrons of the same spin. In this two-particle regime, Ref.~\cite{Hutter2020} shows the universality of the single generalized Slater determinant. Indeed, a generic antisymmetry of two particles can be exactly written as a single two-particle determinant with an appropriately general backflow, and so the architectures compared here are functionally equivalent in terms of their representation power. We see good agreement between the nodal surface cross-sections as early as epoch 2500 (Figure \ref{fig:nodal_surfaces_li}).

However, in beryllium, we observe qualitative differences between the nodal surface cross-sections between the different architectures. If we choose random locations for the three fixed electron positions, we find that the nodal surface cross-sections look much like the smooth spheres in the lithium figure for all architectures. However, in Figure \ref{fig:nodal_surfaces_be} we choose the two opposite-spin electrons to be placed at $(0, \pm 2, 0)$, and we see that the nodal surface cross-sections for FermiNet and FA-1 (Figure \ref{fig:nodal_surfaces_be}) appear to be the union of two smooth surfaces. We were able to confirm that these two surfaces originate from the product structure of the pseudospin terms by removing a psuedospin term and replotting the resulting nodal surface. On the other hand, the nodal surface cross-section obtained from GA and the full single-determinant FermiNet appear to consist of only one smooth surface. This difference aligns with our assertion that the product structure of FermiNet and FA-1 may limit their ability to represent the true nodal surface of the ground state wavefunction. A video is available with rotating views of the final cross-sections for all four wavefunction ansatzes~\footnote{\url{https://youtu.be/67SQXEUCYyY}}.

Our study of the nodal surfaces in this section is importantly limited by the fact that we can only observe a 3-dimensional cross-section of the full nodal surface, so we are not able to directly draw conclusions about the global structure of the nodal surface when using only a single cross-section. Ideally, we could benchmark these plots against the ground truth of the nodal surface generated by the FCI wavefunction for this system. We found, however, that even the qualitative shape of the FCI nodal surface (not depicted here) depends strongly on the choice of the finite sized basis set, and is thus difficult to compare systematically to the VMC-derived wavefunctions.

\subsection{Nitrogen molecule: Performance of the full determinant FermiNet}

\begin{figure}
    \centering
    \includegraphics[width=0.6\textwidth]{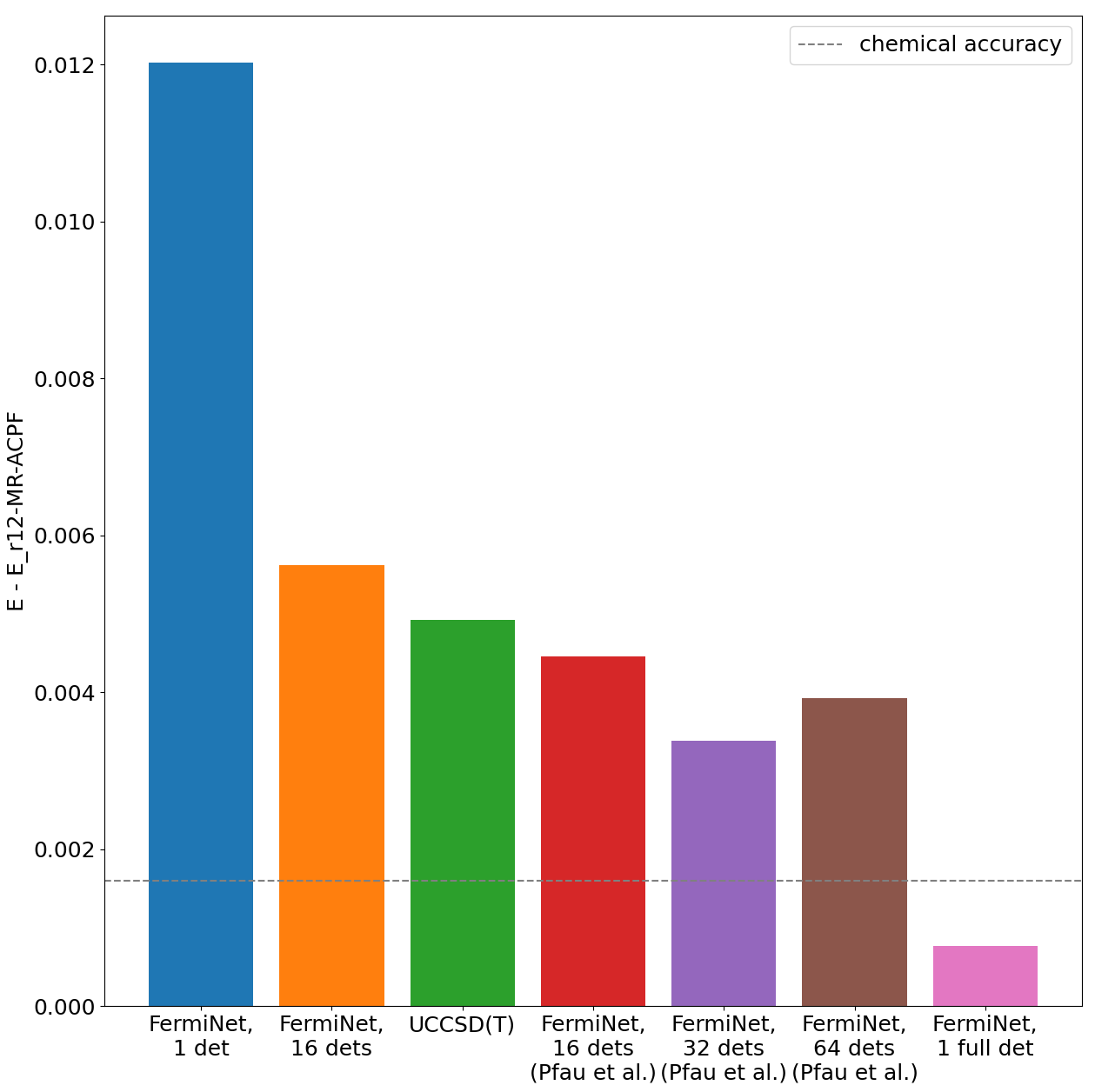}
    \caption{Visual comparison of different FermiNet architectures on the nitrogen molecule with bond length stretched to 4.0 Bohr. Energies shown are the absolute energy difference (in Hartrees) between the attained energy and the r$_{12}$-MR-ACPF computational reference \cite{Gdanitz1998}. The UCCSD(T) and the Pfau et al. energies were extracted from Figure 5 in reference \cite{Pfau2020}. The dashed line indicates chemical accuracy.}
    \label{fig:stretched_n2_compare}
\end{figure}

We finally provide a comparison of a few different FermiNet architectures on the stretched nitrogen molecule, which is a challenging strongly correlated system. Pfau et al. \cite{Pfau2020} demonstrate that the standard FermiNet architecture is not able to accurately model this system, particularly around the dissociating bond length of 4.0 Bohr. We corroborate this finding on the FermiNet repository using 1 and 16 determinants, with results reported in Figure~\ref{fig:stretched_n2_compare}. The energies may be found in \cref{section:stretched_n2_compare}. The performance of the standard multiple-determinant FermiNet is improved somewhat over the standard single-determinant FermiNet, but the energy does not approach the best available computational benchmark obtained by the r$_{12}$-MR-ACPF method \cite{Gdanitz1998}. We are unable to test the FA and GA architectures on this system within the constraints of our computational resources. However, we find that the full single-determinant FermiNet is able to outperform the standard 64-determinant FermiNet and come within chemical accuracy of the benchmark value. We observed a non-trivial amount of run-to-run variance in our results, indicating that initialization and optimization methods for FermiNet-like architectures may require further investigation. Nonetheless, we were able to replicate this result on several distinct optimization runs, and we verified the energies obtained by converting the parameters to the form required by our VMCNet repository and doing a pure MCMC evaluation with rigorous estimates of the statistical error. This astonishing result implies the need for further exploration into the potential universality properties of the full single-determinant FermiNet for strongly correlated problems in quantum chemistry.

\section{Discussion}
We show that explicitly antisymmetrized neural networks can be used to advance the understanding of the performance of neural network based VMC ansatzes.
By replacing the antisymmetric layer in the FermiNet with a generic antisymmetrized neural network with relatively few nodes and a simple Jastrow factor, we find that the resulting FermiNet-GA structure is highly expressive and can yield accurate ground state energies (error of the correlation energy is less than $1\%$). 
On the other hand, if we replace each individual pseudospin determinant of FermiNet with an explicitly antisymmetrized neural network, the resulting FermiNet-FA-$K$ structure does not outperform the $K$-determinant FermiNet.
These observations suggest that the lack of expressiveness of the standard single-determinant FermiNet structure may be largely due to the product structure of the two pseudospin components of its determinant layer.
This motivates us to investigate the ``full determinant''  mode of the FermiNet, which significantly improves the accuracy compared to both standard single-determinant FermiNet and FermiNet-FA-1.
We observed a significant amount of run-to-run variance between training runs with identical architectures, indicating that further understanding of initialization and optimization techniques is needed.

One of our original motivations for developing FermiNet-GA was to resolve the challenges of the stretched nitrogen molecule as discussed in Ref.~\cite{Pfau2020}.
This is a challenging strongly correlated chemical system, and the use of a multiple FermiNet determinant structure is still insufficient compared to the best computational results available.
We were unable to apply the GA layer to this system yet due to its prohibitive computational cost.
However, inspired by the success of the full determinant on smaller systems, we investigate its performance on the nitrogen molecule around a challenging bond length of 4.0 Bohr. We find that the full single-determinant FermiNet achieves an energy within 0.4 kcal/mol of the r$_{12}$-MR-ACPF method \cite{Gdanitz1998}, which is currently the best available computational benchmark.

As part of our work, we contribute a flexible, modular variational Monte Carlo repository called VMCNet~\cite{vmcnet2021github}, built on the JAX machine learning framework \cite{jax2018github}. VMCNet is inspired by the JAX branch of the FermiNet repository \cite{Spencer2020} but uses the Flax API to facilitate developing new components and experimenting with different components of various architectures. We keep the model construction code in a separate submodule from the code for training, sampling, and evaluation, and we leverage simple asynchronous logging to enable monitoring of the training process. This work focuses on VMC simulation in the first quantization, but VMCNet can also be extended to simulate quantum systems in a second quantized form~\cite{NegeleOrland1988} as well. 

We hope this work provides a first step towards understanding the expressiveness of these FermiNet-like architectures. Our results suggest the utility of diagnostic tools such as explicitly antisymmetrized neural networks for building such understanding. They also suggest the need to further explore the potential universality properties of the full determinant mode of FermiNet. 
By further improvement upon the architecture and the optimization of the full determinant FermiNet, 
in the best case, it may be possible to consistently achieve accurate results for a large class of physical and chemical systems of interest.

\section*{Acknowledgements}
This work was partially supported by the Air Force Office of Scientific Research under award number FA9550-18-1-0095 (J.L.),  by the NSF under Grant No. DMS-1652330 (G.G.), and by the Department of Energy under Grant No. DE-SC0017867  and the CAMERA program  (L.L.). 
L.L. is a Simons Investigator.
We would like to acknowledge the use of computational resources at the Berkeley Research Computing (BRC) program at the University of California, Berkeley, the Google Cloud Platform (GCP), and the National Energy Research Scientific Computing Center (NERSC).
We thank Giuseppe Carleo, Bryan Clark, Di Luo, James Stokes, Jiefu Zhang, Xiaojie Wu, and Fabian Faulstich for their helpful discussions. We also thank the hospitality of the American Institute of
Mathematics (AIM) for the SQuaREs program ``Deep learning and quantum Monte Carlo'' in 2021.
\appendix
\section{Hyperparameters}
\label{section:hyperparams}
In Table \ref{tab:hyperparams} we list the hyperparameters used in our runs for the KFAC optimizer. For the stretched N$_2$ geometry, to replicate the results reported by Ref.~\cite{Pfau2020} as closely as possible, 4000 walkers were used instead of 2000, a two-electron stream width of 32 was used instead of 16, and when it resulted in lower energies, pretraining was also used for 1000 iterations.

\begin{table}
        \caption{Table of hyperparameters for KFAC used during training.}
        \label{tab:hyperparams}
        \begin{tabular}{@{}cc@{}}
        \toprule
        Hyperparameter                                         & Value                            \\ \midrule
        Dense nodes per layer in antisymmetrized part          & 64                               \\
        Layers per ResNet in antisymmetrized part              & 2                                \\
        One-electron stream width                              & 256                              \\
        Two-electron stream width                              & 16                               \\
        Number of layers in equivariant part                   & 4 \\
        Kernel initializers for dense layers                   & orthogonal                       \\
        Bias initializers for dense layers                     & random normal                    \\
        Backflow activation function                           & tanh                             \\
        ResNet antisymmetry activation function                & tanh                             \\
        Jastrow (backflow) activation function                 & gelu                             \\
        Number of walkers                                      & 2000                             \\
        Learning rate                                          & $5\cdot 10^{-2} / (1 + 10^{-4}t)$ \\
        Optimizer                                              & KFAC                             \\
        Threshold constant for local energy clipping           & 5.0                              \\
        MCMC steps between updates                             & 10                               \\
        Training iterations (number of parameter updates)      & 2e5                              \\ 
        Evaluation iterations (samples collected every 10)     & 2e5                              \\\bottomrule
        \end{tabular}
\end{table}

\section{Code benchmarking}
\label{section:code_benchmarking}

To demonstrate that our results for the original FermiNet are comparable to those reported by \cite{Pfau2020, Spencer2020}, we show that results obtained using the VMCNet repository are quantitatively comparable to that of the JAX branch of the FermiNet repository presented in \cite{Spencer2020} on several small systems. We compare the behavior on both the nitrogen atom and the square H$_4$ model (Figure \ref{fig:H4_setup}), using settings corresponding to the original FermiNet model in both repositories. For the nitrogen atom, we compare results with 1, 2, and 4 determinants, while for the H$_4$ square we compare results with just 1 and 2 determinants, since 2 determinants already captures essentially 100\% of the correlation energy. All results presented here come from our own numerical experiments with either the VMCNet repository or the publicly available JAX branch of the FermiNet repository. Since VMCNet does not support Hartree-Fock based pretraining, we turned this feature off in the FermiNet repository to make the comparison fair. Turning off pretraining reduces the consistency of the FermiNet optimization on some systems. In particular, when using multiple determinants for the nitrogen atom, we found that some runs both of our own code and of the FermiNet code without pretraining get stuck in local minima and never reach the lowest energy possible. This phenomenon may merit further investigation. For now, to account for this run-to-run variance, we have taken the best of several runs for all multi-determinant experiments on the nitrogen atom. 

\begin{figure}
\centering
    \begin{subfigure}[b]{0.45\textwidth}
        \includegraphics[width=\textwidth]{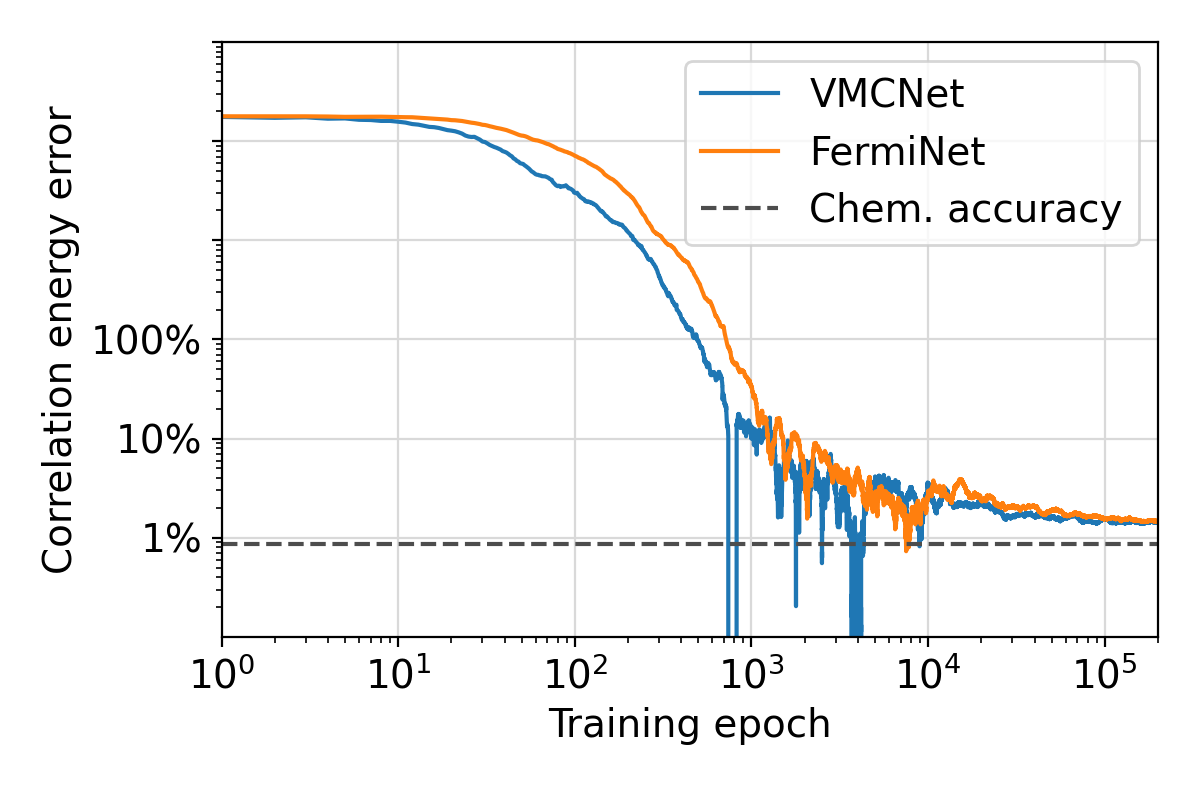}
        \caption{1 determinant}
    \end{subfigure}
    \begin{subfigure}[b]{0.45\textwidth}
        \includegraphics[width=\textwidth]{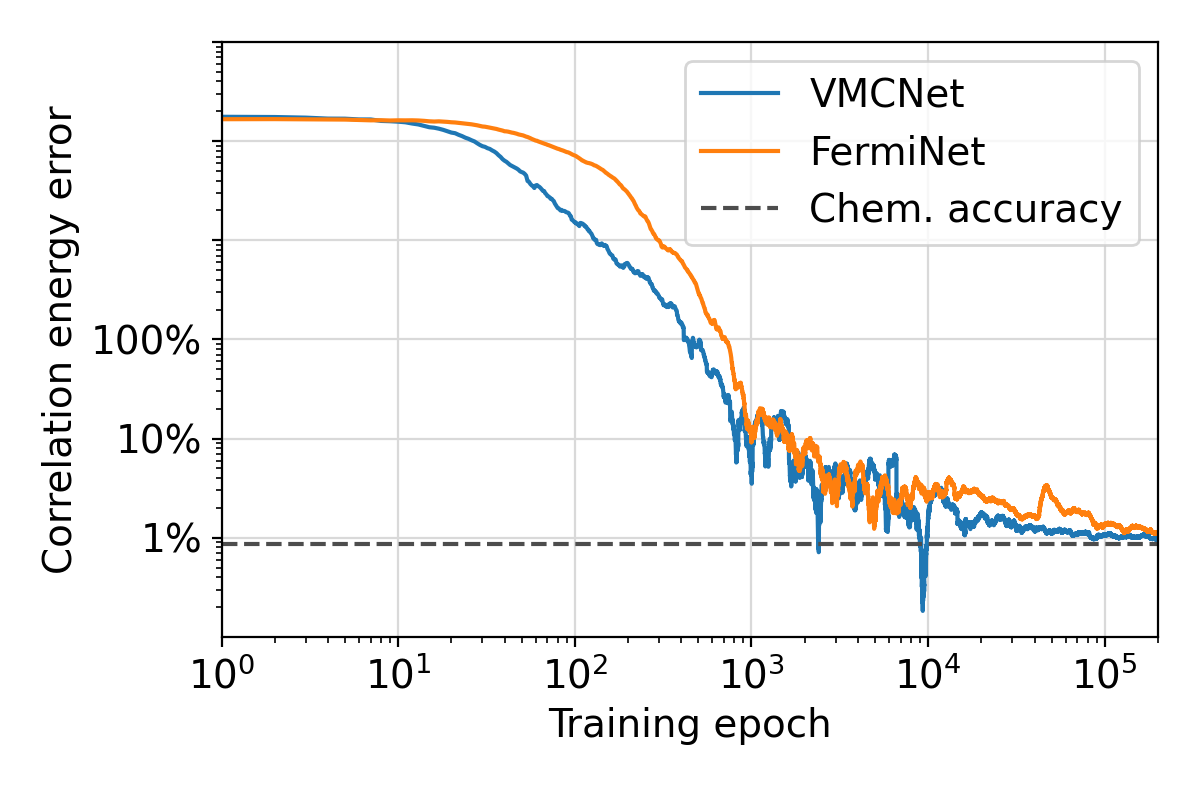}
        \caption{2 determinants}
    \end{subfigure}
    \begin{subfigure}[b]{0.45\textwidth}
        \includegraphics[width=\textwidth]{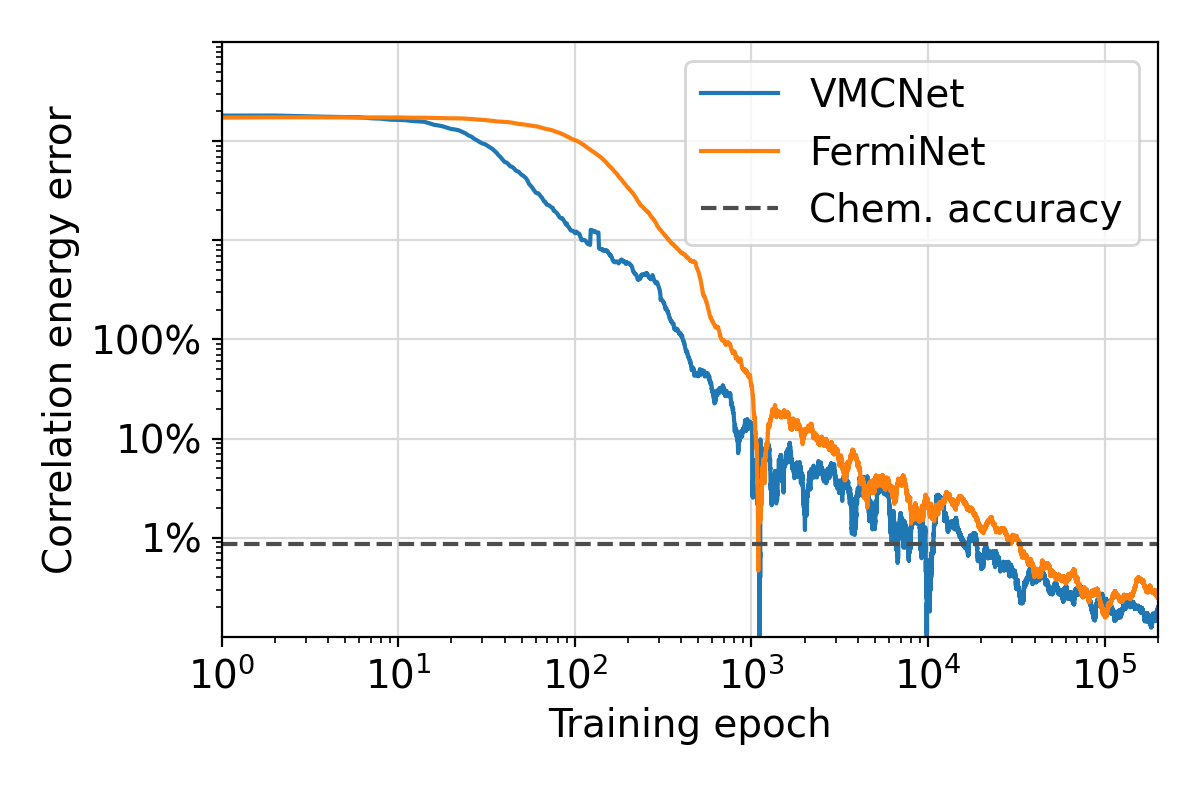}
        \caption{4 determinants}
    \end{subfigure}
    \caption{Training runs on VMCNet and FermiNet repositories with the FermiNet architecture on the nitrogen atom. At each epoch, rolling averages of the previous 10\% of training epochs are shown here for clarity. One epoch means one parameter update. Data for 2 and 4 determinants represent the best of several runs to account for run-to-run variance observed in both repositories.}
    \label{fig:vmc_fnt_N}
\end{figure}

\begin{figure}
\centering
    \begin{subfigure}[b]{0.45\textwidth}
        \includegraphics[width=\textwidth]{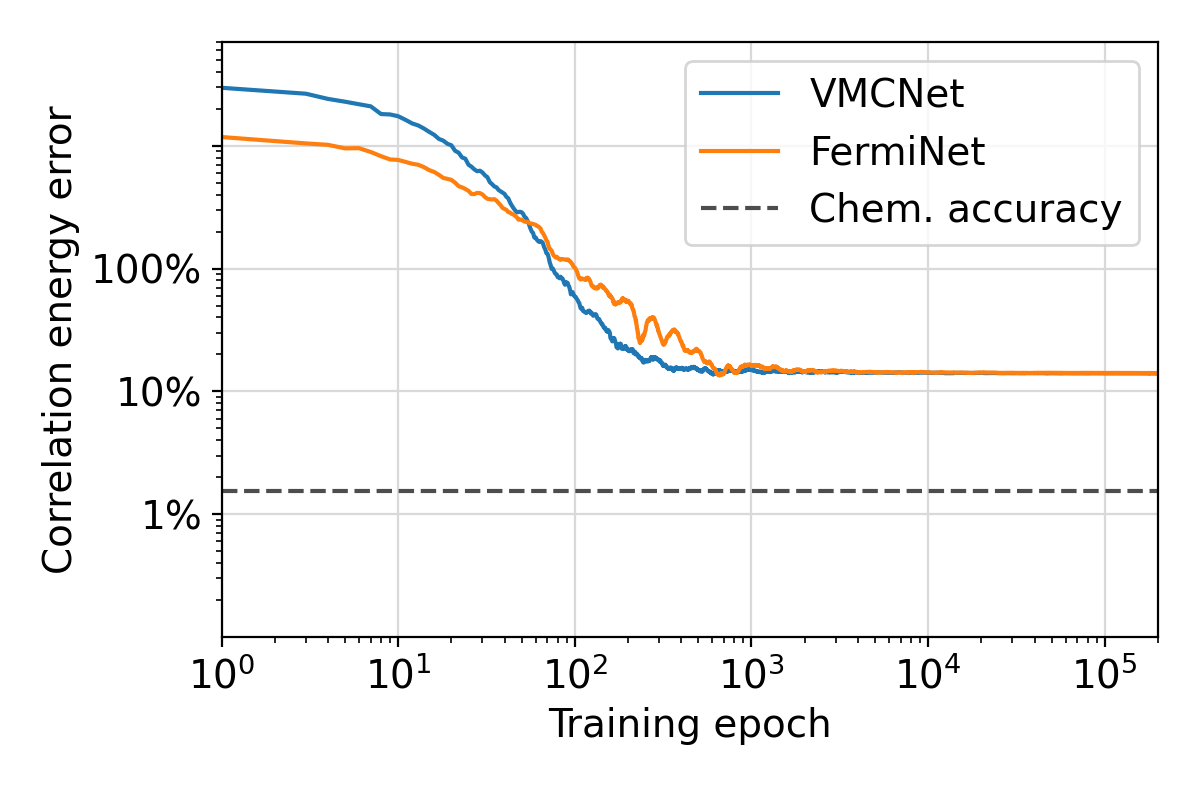}
        \caption{1 determinant}
    \end{subfigure}
    \begin{subfigure}[b]{0.45\textwidth}
        \includegraphics[width=\textwidth]{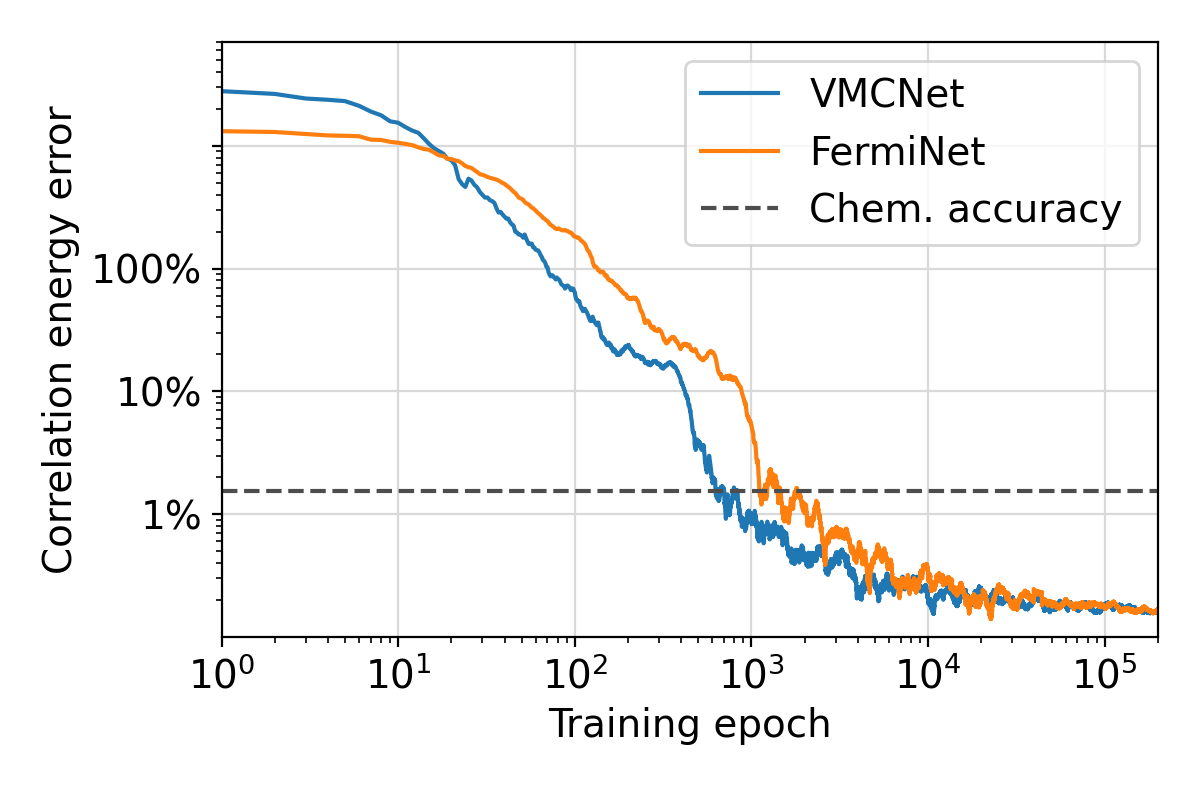}
        \caption{2 determinants}
    \end{subfigure}
    \caption{Training runs on VMCNet and FermiNet repositories with the FermiNet architecture on the H4 square. At each epoch, rolling averages of the previous 10\% of training epochs are shown here for clarity. One epoch means one parameter update.}
    \label{fig:vmc_fnt_H4}
\end{figure}

Representative training graphs can be found for the nitrogen atom in Figure \ref{fig:vmc_fnt_N} and for the H$_4$ square in Figure \ref{fig:vmc_fnt_H4}. The values of the final energies obtained are presented in Tables \ref{tab:vmc_fnt_N_compare} and \ref{tab:vmc_fnt_H4_compare}, respectively. On both systems, the results of VMCNet are approximately equivalent to the results of FermiNet. The two repositories behave somewhat differently in the first 1,000 epochs of training, with VMCNet often optimizing more quickly in this regime. However, the optimization trajectories are largely indistinguishable by 10,000 epochs and the final energies achieved are within a small margin of error of each other in all cases.

\begin{table}
\centering
\caption{Comparison of the VMCNet repository with the FermiNet repository on the nitrogen atom, with 1, 2, and 4 determinants. Data for 2 and 4 determinants represent the best of several runs to account for run-to-run variance observed in both repositories.}
\label{tab:vmc_fnt_N_compare}
\begin{tabular}{ccccccc}
\toprule
Repository  & 1 det & corr \% & 2 det & corr \% & 4 det & corr \%  \\ \midrule
VMCNet   & -54.5864(1)  & 98.48(8)\% & -54.58739(4) & 99.02(2)\% & -54.58891(4) & 99.85(2)\% \\
FermiNet & -54.58654(5) & 98.56(3)\% & -54.58711(6) & 98.87(3)\% & -54.58870(4) & 99.73(4)\%  \\ \bottomrule
\end{tabular}
\end{table}

\begin{table}
\centering
\caption{Comparison of the VMCNet repository with the FermiNet repository on the H$_4$ square, with 1 and 2 determinants.}
\label{tab:vmc_fnt_H4_compare}
\begin{tabular}{ccc}
\toprule
Repository         & 1 det & 2 det  \\ \midrule
VMCNet & -1.424531(7) & -1.438804(5) \\
FermiNet & -1.424429(7) & -1.438796(5) \\ \bottomrule
\end{tabular}
\end{table}

\section{Numerical stability and computational cost of the antisymmetric layer}\label{sec:challenges}

\begin{figure}
\centering
\includegraphics[width=0.5\textwidth]{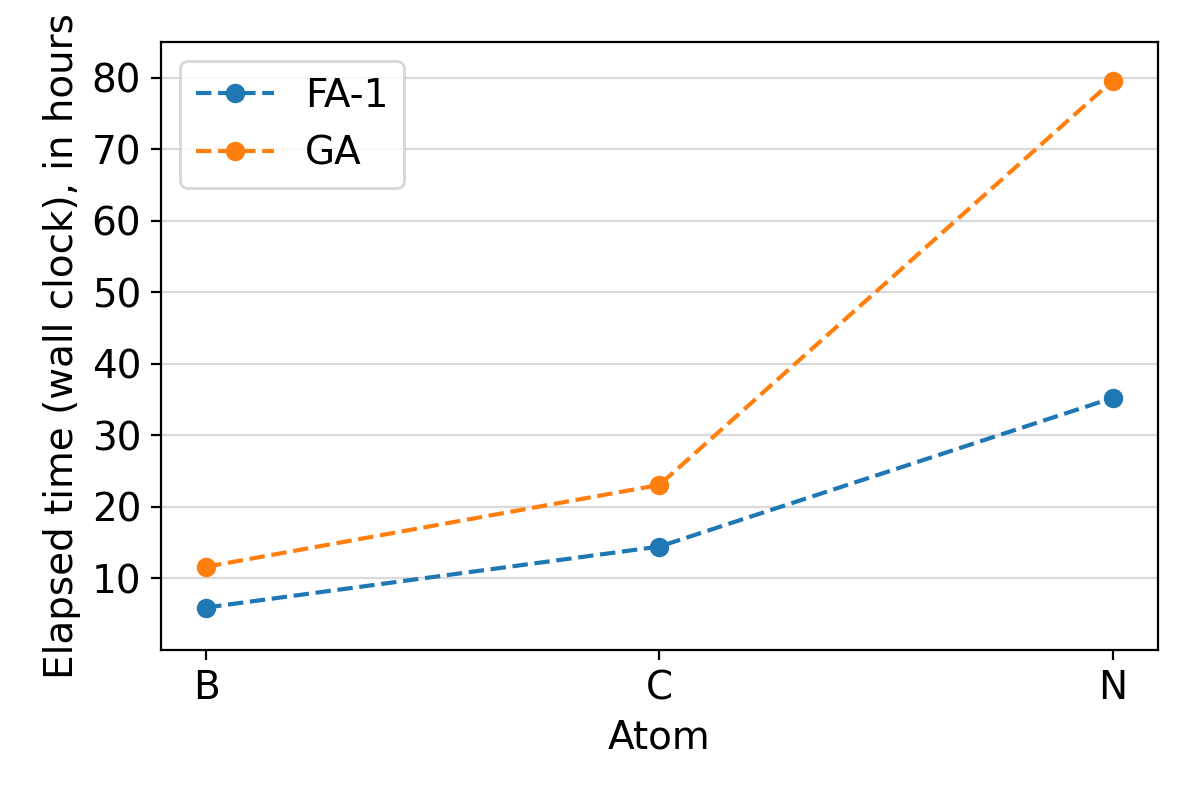}
\caption{Wall clock hours elapsed during training for the generic antisymmetry and the  factorized antisymmetrized of rank 1 with the backflow-based Jastrow for the boron, carbon, and nitrogen atoms. The training was performed for $2\times 10^5$ epochs on 2 A100 GPUs.}
\label{fig:a100_timings}
\end{figure}

One challenge we faced when training the generic antisymmetric architecture was the numerical sign cancellation near the nodal hypersurface. When computing FermiNet-GA in single precision, we invariably encounted NaNs (not-a-number). Some investigation revealed that the computation of $\Psi$ could yield slightly different results depending on whether it was calculated during a simple forward pass evaluation or a gradient calculation involving a forward and backward pass. Due to this numerical inconsistency, the Metropolis-Hastings procedure would sometimes sample points on or extremely close to the nodal hypersurface. To contend with this in our experiments, we used double precision end-to-end, i.e. converted all arrays to double precision. It is possible that a more efficient implementation might use double precision only in the antisymmetric layer or only when evaluating the local energy. The use of double precision led us to use A100 GPUs, which have significantly better performance for these higher precision calculations than consumer GTX GPUs. We used GTX 2080TI GPUs for our experiments which did not require double precision. Despite these powerful GPUs, the unfavorable scaling of the brute-force antisymmetry meant that we reached the limits of our group's resources with the calculations on the oxygen atom. On 4 A100 GPUs, training the FermiNet-GA architecture with the simplified Jastrow on the oxygen atom took 137 hours. In Figure \ref{fig:a100_timings}, we show the wall clock time used to train the generic and factorized antisymmetric architectures for boron through nitrogen.

\section{Basis set extrapolation for the square H$_4$ model}
\label{sec:basis_set_extrap}
The Hartree-Fock (HF) and full configuration interaction (FCI) values for the square H$_4$ model were extrapolated to the complete basis limit using cc-pvXz basis sets using PySCF~\cite{Sun2018,Sun2020}. For completeness, we reproduce the details of the extrapolation here.

The complete basis set Hartree-Fock energies were obtained by a fit to the function
\begin{equation}
E_{\mathrm{HF}}(X) = E_{\mathrm{HF}}(\mathrm{CBS}) + a\exp(-bX),
\end{equation}
where $E_{\mathrm{HF}}(X)$ is the Hartree-Fock energy computed with cc-pvXz and the parameters $E_{\mathrm{HF}}(\mathrm{CBS})$, $a$, and $b$ are determined with a non-linear least-squares fit. Similarly, the complete basis set correlation energies are obtained by a fit to the function
\begin{equation}
E_{\mathrm{corr}}(X) = E_{\mathrm{corr}}(\mathrm{CBS}) + aX^{-3},
\end{equation}
where $E_{\mathrm{corr}}(X)$ is the difference between the FCI and Hartree-Fock energies on the cc-pvXz basis and the parameters $E_{\mathrm{corr}}(\mathrm{CBS})$ and $a$ are determined with a non-linear least-squares fit.

For RHF/UHF at bond length 1.0, we used $X = 2, 3, 4, 5$, as the orbital overlap matrix became too ill-conditioned for larger $X$. For RHF/UHF at bond length 4.0, we used $\mathrm{X} = 5, 6, 8$. The FCI calculations were done using restricted Hartree-Fock (RHF) as the initial reference, and the correlation energy was computed as the difference between the FCI and RHF energies. For the extrapolation of the RHF-FCI correlation energy we used $\mathrm{X} = 3, 4$. Due to the relative unreliability of the data points from the small double-zeta basis set and the cost of the quintuple-zeta basis set, these points were not included in the extrapolation for the correlation energy. Judging simply from the square root of the variance of the parameter fit, the basis set extrapolation error is at least two orders of magnitude larger than that of the Monte Carlo error in the estimates of the VMC-derived energies, so fewer significant digits are reported for the RHF/UHF/FCI results. 

\section{Gradient calculation}
\label{section:gradient}
In this section we derive an unbiased estimate for the gradient of the expected energy of the wavefunction. Recall that the expected energy is given by
\begin{equation}
\loss = \frac{ \intR{E_L(\vR;\theta)\abs{\Psi_{\theta}}^2 }}{\intR{ \abs{\Psi_{\theta}}^2}}.
\end{equation}
For the purposes of the following derivation, we will let $\theta$ be a real number representing any parameter. The derivative of the integrand in the denominator with respect to $\theta$ is
\begin{equation}
\partial_\theta \abs{\Psi_{\theta}}^2
= (\partial_\theta\Psi_{\theta})\Psi_{\theta}^* + \Psi_{\theta}(\partial_\theta\Psi_{\theta}^*)
= \left(\frac{\partial_\theta\Psi_{\theta}}{\Psi_{\theta}} + \frac{\partial_\theta\Psi_{\theta}^*}{\Psi_{\theta}^*}\right)\abs{\Psi_{\theta}}^2,
\end{equation}
and the derivative of the integrand in the numerator with respect to $\theta$ is
\begin{equation}
\begin{split}
\partial_\theta\left[E_L(\vR;\theta)\abs{\Psi_{\theta}}^2\right] &= \partial_\theta\left[\Psi_{\theta}^*H\Psi_{\theta}\right]\\
&= \partial_\theta\Psi_{\theta}^*H\Psi_{\theta} + \Psi_{\theta}^*H\partial_\theta\Psi_{\theta} \\
&= \frac{\partial_\theta\Psi_{\theta}^*}{\Psi_{\theta}^*}E_L(\vR;\theta)\abs{\Psi_{\theta}}^2 + \Psi_{\theta}^*H\partial_\theta\Psi_{\theta}
\end{split}
\end{equation}
To treat the latter term in the last expression above, we also take advantage of the following identity, which uses the essential self-adjointness of the Born-Oppenheimer Hamiltonian $H$ \cite{Reed1975}:
\begin{equation}
\intR{\Psi_{\theta}^*(H\partial_\theta\Psi_{\theta})} = \intR{\partial_\theta\Psi_{\theta}(H\Psi_{\theta}^*)} = \intR{\frac{\partial_\theta\Psi_{\theta}}{\Psi_{\theta}}E_L(\vR;\theta)^*\abs{\Psi_{\theta}}^2}
\end{equation}
Using these facts, the following calculation gives the derivative of the expected energy $\mathcal{L}$ with respect to $\theta$:
\begin{equation}
\begin{split}
\partial_\theta \loss &= \partial_\theta \frac{\intR{ E_L(\vR;\theta)\abs{\Psi_{\theta}}^2} }{\intR{ \abs{\Psi_{\theta}}^2}} \\
&= \frac{\partial_\theta \intR{ E_L(\vR;\theta)\abs{\Psi_{\theta}}^2} }{\intR{ \abs{\Psi_{\theta}}^2}} - \frac{\intR{ E_L(\vR;\theta)\abs{\Psi_{\theta}}^2} }{\intR{ \abs{\Psi_{\theta}}^2}}\frac{\partial_\theta\intR{ \abs{\Psi_{\theta}}^2}}{\intR{ \abs{\Psi_{\theta}}^2}} \\
&= \frac{\intR{((\partial_\theta\Psi_{\theta}^* / \Psi_{\theta}^*)E_L(\vR;\theta) + (\partial_\theta\Psi_{\theta} / \Psi_{\theta}) E_L(\vR;\theta)^*)\abs{\Psi_{\theta}}^2}}{\intR{ \abs{\Psi_{\theta}}^2}} \\
&\quad- \frac{\intR{E_L(\vR;\theta) \abs{\Psi_{\theta}}^2}}{\intR{ \abs{\Psi_{\theta}}^2}}\frac{\intR{(\partial_\theta\Psi_{\theta}^* / \Psi_{\theta}^* + \partial_\theta\Psi_{\theta} / \Psi_{\theta})\abs{\Psi_{\theta}}^2}}{\intR{ \abs{\Psi_{\theta}}^2}}\\
&=\frac{\intR{[(\partial_\theta\Psi_{\theta}^* / \Psi_{\theta}^*)(E_L(\vR;\theta) - \mathcal{L}(\theta)) + (\partial_\theta\Psi_{\theta} / \Psi_{\theta})(E_L(\vR;\theta)^* - \mathcal{L}(\theta))]\abs{\Psi_{\theta}}^2}}{\intR{ \abs{\Psi_{\theta}}^2}}.
\end{split}
\end{equation}
When the wavefunction is real, we may simplify this to
\begin{equation}
\partial_\theta \loss = \frac{\intR{2(\partial_\theta\Psi_{\theta} / \Psi_{\theta})(E_L(\vR;\theta) - \mathcal{L}(\theta)) \abs{\Psi_{\theta}}^2}}{\intR{ \abs{\Psi_{\theta}}^2}}.
\end{equation}
We may then use Monte Carlo sampling to estimate the gradient as
\begin{equation}
\partial_\theta\mathcal{L}(\theta) \approx \frac{1}{|\xi_{\theta}|}\sum_{\vR\in\xi_\theta} 2(\partial_\theta\log\abs{\Psi_\theta})(E_L(\vR;\theta) - \tilde{\mathcal{L}}(\theta))
\end{equation}
where $\xi_\theta$ are a set of samples from the density $p_\theta(\vR) = |\Psi_\theta(\vR)|^2 / \intR{ \abs{\Psi_{\theta}(\vR)}^2}$.

\section{Sampling and gradient clipping}
\label{section:sampling_and_clipping}
We use the Metropolis-Hastings algorithm to sample electron configurations from the distribution defined by $\Psi(\vec{X})$. We use a gaussian proposal function with an isotropic step width, which we dynamically update throughout the optimization in order to keep the average acceptance ratio near a target value, for which we use $0.5$. We maintain this ratio through a simple scheme that increases the step width by a small amount if the acceptance ratio strays too far above the target, and similarly decreases it by a small amount if the ratio strays too far below the target. We perform such updates every 100 moves, averaging the acceptance ratio over the previous hundred steps in order to avoid overzealously updating the step width due to noise in the acceptance ratio.

In order to reduce the amount of correlation between the samples used for subsequent parameter updates, we take $10$ walker steps between each gradient calculation and parameter update. While skipping steps theoretically does not produce a higher effective sample size than simply using every step, it is practically beneficial to skip steps because the local energy calculation required for a parameter update is significantly more computationally expensive than the wave function amplitude calculation required for each move. This means we can take a number of intermediate steps in order to produce significantly less correlated samples with a small computational overhead.

As is common in quantum Monte Carlo \cite{Umrigar1993}, in order to reduce the noise in the training process, we additionally clip the local energies calculated in each batch of samples to be closer to some estimator of the energy intended to reduce the effect of outliers in the gradient. Specifically, given a batch of local energies $E_1, E_2, \ldots ,E_n$, we calculate the median local energy $E_M$ and then calculate the average deviation from the median (total variation) as 
\begin{equation}
TV = \frac{1}{n} \sum_{i} \abs{E_i - E_M}. 
\end{equation}
We then replace $E_i$ with $E_M$ whenever $\abs{E_i - E_M} > 5\cdot TV$. In practice, we have found that this produces a less noisy and more effective optimization process than including all of the unclipped local energies. During the final Monte Carlo evaluation of the energy after training, no local energy clipping is performed in order to avoid bias in the energy estimate.

\section{Factorized antisymmetry versus FermiNet}

We record the numerical results comparing FermiNet-FA-$K$ to the standard $K$-determinant FermiNet in Table~\ref{tab:vrank_k}, also shown in Figure~\ref{fig:rank_k}. Due to the limits of our computational resources, we did not compute the $K = 3, 4$ results for FA-$K$ with the backflow-based Jastrow.

\begin{table}
\centering
\caption{Numerical results for comparison of factorized antisymmetry of rank $K$ with $K$-determinant FermiNet for $K=1,2,3,4$ on the nitrogen atom.}
\label{tab:vrank_k}
\begin{tabular}{ccccccc}
\toprule
& FA-$K$, one-body Jastrow & corr\% & FA-$K$, backflow Jastrow & corr\% & $K$-det. FermiNet & corr\% \\ \midrule
$K$=1 & 54.58637(4) & 98.47(2) & -54.58664(4) & 98.61(2) & -54.5864(1)  & 98.48(8)  \\
$K$=2 & -54.58715(4) & 98.89(2) & -54.58733(4) & 98.99(2) & -54.58739(4) & 99.02(2) \\
$K$=3 & -54.58803(3) & 99.37(1) &  --    &   --   & -54.58817(4) & 99.44(2) \\
$K$=4 & -54.58855(5) & 99.65(3) &  --      &  --    & -54.58891(4) & 99.85(2) \\
\bottomrule
\end{tabular}
\end{table}

\section{Stretched nitrogen molecule}
\label{section:stretched_n2_compare}
We record the numerical results comparing various FermiNet architectures on the nitrogen molecule with bond length 4.0 Bohr in Table~\ref{tab:stretched_n2_compare}. These results are also depicted in Figure~\ref{fig:stretched_n2_compare}. As noted in \cref{section:hyperparams}, we use 4000 walkers and used runs both with and without pretraining on the FermiNet repository in order to replicate the previously reported results in \cite{Pfau2020} as closely as possible. The parameters were reloaded and evaluated using the VMCNet repository to ensure reproducibility and to evaluate the quality of the sampling procedure. The UCCSD(T) and Pfau et al. results were extracted from Figure 5 in \cite{Pfau2020}.

\begin{table}
\centering
\caption{Comparison of FermiNet architectures on the nitrogen molecule with bond length stretched to 4.0. The experimental result is computed from the $\mathrm{MLR}_4(6,8)$ fitted potential curve recommended by the authors of \cite{LeRoy2006}.}
\label{tab:stretched_n2_compare}
\begin{tabular}{@{}lc@{}}
\toprule
                                        & energy    \\ \midrule
RHF                                     & -108.3101 \\
\midrule
FermiNet, 1 det                         & -109.1827 \\
FermiNet, 16 dets                       & -109.1891 \\
FermiNet, 16 dets (Pfau et al. \cite{Pfau2020})                       & -109.1903 \\
FermiNet, 32 dets (Pfau et al. \cite{Pfau2020})                       & -109.1913 \\
FermiNet, 64 dets (Pfau et al. \cite{Pfau2020})                       & -109.1908 \\
FermiNet, 1 full det                    & -109.1940 \\
\midrule
UCCSD(T) (Pfau et al. \cite{Pfau2020})                & -109.1898 \\
r$_{12}$-MR-ACPF \cite{Gdanitz1998}     & -109.1947 \\
\midrule
experiment \cite{LeRoy2006}             & -109.2021 \\\bottomrule
\end{tabular}
\end{table}

\bibliography{refs}

\end{document}